\documentclass[longauth]{aa} 

\usepackage{graphicx}
\usepackage{amsmath}
\usepackage{longtable}
\begin{document}

\title{The changing face of AU Mic~b: stellar spots, spin-orbit commensurability, and transit timing variations as seen by CHEOPS and TESS}

\titlerunning{The changing face of AU Mic~b}

\author{
Gy.M.~Szab\'o\inst{1,2}
    \and
D.~Gandolfi\inst{3}
    \and
A.~Brandeker\inst{4}
    \and
Sz.~Csizmadia\inst{5}
    \and
Z.~Garai\inst{1,2,6}
    \and
N.~Billot\inst{7}
    \and
C.~Broeg\inst{8}
    \and
D.~Ehrenreich\inst{7}
    \and
A.~Fortier\inst{8}
    \and
L.~Fossati\inst{9}
    \and
S.~Hoyer\inst{10}
    \and
L.~Kiss\inst{11,12,13}
    \and
A.~Lecavelier~des~Etangs\inst{14}
    \and
P.F.L.~Maxted\inst{15}
    \and
I.~Ribas\inst{16,17}
    \and
Y.~Alibert\inst{8}
    \and
R.~Alonso\inst{18, 19}
    \and
G.~Anglada~Escud\'e\inst{}
    \and
T.~B\'arczy\inst{20}
    \and
S.C.C.~Barros\inst{21,22}
    \and
D.~Barrado\inst{23}
    \and
W.~Baumjohann\inst{9}
    \and
M.~Beck\inst{7}
    \and
T.~Beck\inst{24}
    \and
A.~Bekkelien\inst{7}
    \and
X.~Bonfils\inst{25}
    \and
W.~Benz\inst{8, 24}
    \and
L.~Borsato\inst{26}
    \and
M-D.~Busch\inst{8}
    \and
J.~Cabrera\inst{5}
    \and
S.~Charnoz\inst{27}
    \and
A.~Collier~Cameron\inst{28}
    \and
C.~Corral~Van~Damme\inst{29}
    \and
M.B.~Davies\inst{30}
    \and
L.~Delrez\inst{31, 32}
    \and
M.~Deleuil\inst{10}
    \and
O.D.S.~Demangeon\inst{10,21}
    \and
B.-O.~Demory\inst{24}
    \and
A.~Erikson\inst{5}
    \and
M.~Fridlund\inst{33, 34}
    \and
D.~Futyan\inst{7}
    \and
A.~Garc\'ia~Mu\~noz\inst{35}
    \and
M.~Gillon\inst{32}
    \and
M.~Guedel\inst{36}
    \and
P.~Guterman\inst{10,37}
    \and
K.~Heng\inst{24,38}
    \and
K.G.~Isaak\inst{29}
    \and
G.~Lacedelli\inst{39,26}
    \and
J.~Laskar\inst{40}
    \and
M.~Lendl\inst{7}
    \and
C.~Lovis\inst{7}
    \and
A.~Luntzer\inst{36}
    \and
D.~Magrin\inst{26}
    \and
V.~Nascimbeni\inst{26}
    \and
G.~Olofsson\inst{4}
    \and
H.P.~Osborn\inst{41,42}
    \and
R.~Ottensamer\inst{36}
    \and
I.~Pagano\inst{43}
    \and
E.~Pallé\inst{18,19}
    \and
G.~Peter\inst{44}
    \and
D.~Piazza\inst{8}
    \and
G.~Piotto\inst{39,26}
    \and
D.~Pollacco\inst{38}
    \and
D.~Queloz\inst{7, 45}
    \and
R.~Ragazzoni\inst{39,26}
    \and
N.~Rando\inst{29}
    \and
H.~Rauer\inst{5, 35,46}
    \and
N.C.~Santos\inst{21,22}
    \and
G.~Scandariato\inst{43}
    \and
D.~S\'egransan\inst{7}
    \and
L.M.~Serrano\inst{}
    \and
D.~Sicilia\inst{43}
    \and
A.E.~Simon\inst{8}
    \and
A.M.S.~Smith\inst{5}
    \and
S.G.~Sousa\inst{21}
    \and
M.~Steller\inst{9}
    \and
N.~Thomas\inst{8}
    \and
S.~Udry\inst{7}
    \and
V.~Van~Grootel\inst{31}
    \and
N.A.~Walton\inst{45}
    \and
T.G.~Wilson\inst{28}
}

\authorrunning{Szab\'o et al.}

\institute{
ELTE E\"otv\"os Lor\'and University, Gothard Astrophysical Observatory, 9700 Szombathely, Szent Imre h. u. 112, Hungary \and
MTA-ELTE Exoplanet Research Group, 9700 Szombathely, Szent Imre h. u. 112, Hungary \and
INAF, Osservatorio Astrofisico di Torino, via Osservatorio 20, 10025 Pino Torinese, Italy \and
Department of Astronomy, Stockholm University, AlbaNova University Center, 10691 Stockholm, Sweden \and
Institute of Planetary Research, German Aerospace Center (DLR), Rutherfordstrasse 2, 12489 Berlin, Germany \and
Astronomical Institute, Slovak Academy of Sciences, 05960 Tatransk\'a Lomnica, Slovakia \and
Observatoire de Gen\`eve, Universit\'e de Gen\`eve, Chemin Pegasi, 51 1290 Versoix, Switzerland \and
Physikalisches Institut, University of Bern, Gesellsschaftstrasse 6, 3012 Bern, Switzerland \and
Space Research Institute, Austrian Academy of Sciences, Schmiedlstrasse 6, A-8042 Graz, Austria \and
Aix Marseille Univ, CNRS, CNES, LAM, Marseille, France \and
Konkoly Observatory, Research Centre for Astronomy and Earth Sciences, 1121 Budapest, Konkoly Thege Mikl\'os út 15-17, Hungary \and
ELTE E\"otv\"os Lor\'and University, Institute of Physics, P\'azm\'any P\'eter s\'et\'any 1/A, 1117 Budapest, Hungary \and
Sydney Institute for Astronomy, School of Physics A29, University of Sydney, NSW 2006, Australia \and
Institut d’astrophysique de Paris, UMR7095 CNRS, Universit\'e Pierre \& Marie Curie, 98bis blvd. Arago, 75014 Paris, France \and
Astrophysics Group, Keele University, Staffordshire, ST5 5BG, United Kingdom \and
Institut de Ci\`encies de l'Espai (ICE, CSIC), Campus UAB, C/CanMagrans s/n, 08193 Bellaterra, Spain \and
Institut d’Estudis Espacials de Catalunya (IEEC), Barcelona, Spain \and
Instituto de Astrof\'\i{}sica de Canarias (IAC), 38200 La Laguna, Tenerife, Spain \and
Departamento de Astrof\'\i{}sica, Universidad de La Laguna (ULL), E-38206 La Laguna, Tenerife, Spain \and
Admatis, Miskolc, Hungary \and
Instituto de Astrof\'isica e Ci\^encias do Espa\c{c}o, Universidade do Porto, CAUP, Rua das Estrelas, 4150-762 Porto, Portugal \and
Departamento de F\'isica e Astronomia, Faculdade de Ci\^encias, Universidade do Porto, Rua do Campo Alegre, 4169-007 Porto, Portugal \and
Depto. de Astrof\'\i{}sica, Centro de Astrobiologia (CSIC-INTA), ESAC campus, 28692 Villanueva de la Cãda (Madrid), Spain \and
Center for Space and Habitability, Gesellsschaftstrasse 6, 3012 Bern, Switzerland \and
Universit\'e Grenoble Alpes, CNRS, IPAG, 38000 Grenoble, France \and
INAF, Osservatorio Astronomico di Padova, Vicolo dell'Osservatorio 5, 35122 Padova, Italy \and
Institut de Physique du Globe de Paris (IPGP), 1 rue Jussieu, 75005 Paris, France \and
School of Physics and Astronomy, Physical Science Building, North Haugh, St Andrews, United Kingdom \and
ESTEC, European Space Agency, Keplerlaan 1, 2201 AZ Noordwijk, The Netherlands \and
Lund Observatory, Dept. of Astronomy and Theoreical Physics, Lund University, Box 43, 22100 Lund, Sweden \and
Space sciences, Technologies and Astrophysics Research (STAR) Institute, Universit\'e de Li\`ege, All\'ee du six Ao\^ut 19C, 4000 Li\`ege, Belgium \and
Astrobiology Research Unit, Universit\'e de Li\`ege, All\'ee du six Ao\^ut 19C, 4000 Li\`ege, Belgium 
\newpage
\and
Leiden Observatory, University of Leiden, PO Box 9513, 2300 RA Leiden, The Netherlands \and
Department of Space, Earth and Environment, Chalmers University of Technology, Onsala Space Observatory, 43992 Onsala, Sweden \and
Center for Astronomy and Astrophysics, Technical University Berlin, Hardenberstrasse 36, 10623 Berlin, Germany \and
University of Vienna, Department of Astrophysics, T\"urkenschanzstrasse 17, 1180 Vienna, Austria \and
Division Technique INSU, BP 330, 83507 La Seyne cedex, France \and
Department of Physics, University of Warwick, Gibbet Hill Road, Coventry CV4 7AL, United Kingdom \and
Dipartimento di Fisica e Astronomia "Galileo Galilei", Universià degli Studi di Padova, Vicolo dell'Osservatorio 3, 35122 Padova, Italy \and
IMCEE, UMR8028 CNRS, Observatoire de Paris, PSL Univ., Sorbonne Univ., 77 av. Denfert-Rochereau, 75014 Paris, France \and
NCCR/PlanetS, Centre for Space \& Habitability, University of Bern, Bern, Switzerland \and
Department of Physics and Kavli Institute for Astrophysics and Space Research, Massachusetts Institute of Technology, Cambridge, MA 02139, USA \and
INAF, Osservatorio Astrofisico di Catania, Via S. Sofia 78, 95123 Catania, Italy \and
Institute of Optical Sensor Systems, German Aerospace Center (DLR), Rutherfordstrasse 2, 12489 Berlin, Germany \and
Astrophysics Group, Cavendish Laboratory, J.J. Thomson Avenue, Cambridge CB3 0He, United Kingdom \and
Institut f\"ur Geologische Wissenschaften, Freie Universität Berlin, 12249 Berlin, Germany 
}

\date{Received date / Accepted date }

\abstract{
AU Mic is a young planetary system with a resolved debris disc showing signs of planet formation and two transiting warm Neptunes near mean-motion resonances. Here we analyse three transits of AU\,Mic\,b observed with the  \textit{CHaracterising ExOPlanet Satellite} (\textit{CHEOPS}), supplemented with sector 1 and 27 \textit{Transiting Exoplanet Survey Satellite} (\textit{TESS}) photometry, and the \textit{All-Sky Automated Survey (ASAS)} from the ground. The refined orbital period of AU\,Mic\,b is $8.462995 \pm 0.000003$~d, whereas the stellar rotational period is P$_\mathrm{rot}$\,=\,4.8367\,$\pm$\,0.0006\,d. The two periods indicate a 7:4 spin--orbit commensurability at a precision of 0.1\,\%{}. Therefore, all transits are observed in front of one of the four possible stellar central longitudes. This is strongly supported by the observation that the same complex star-spot pattern is seen in the second and third CHEOPS visits that were separated by four orbits (and seven stellar rotations). Using a bootstrap analysis we find that flares and star spots reduce the accuracy of transit parameters by up to 10\,\%{} in the planet-to-star radius ratio and the accuracy on transit time by 3-4 minutes. Nevertheless, occulted stellar spot features independently confirm the presence of transit timing variations (TTVs) with an amplitude of at least 4 minutes. We find that the outer companion, AU\,Mic\,c, may cause the observed TTVs.
}


\maketitle

\section{Introduction}

Very young planetary systems allow us to study planets that recently formed or are still in the process of formation. One of the youngest stars with a planetary system known to date is AU\,Mic, a $\sim$22~Myr-old \citep{2014MNRAS.445.2169M} M1 star belonging to the $\beta$~Pic moving group \citep{2006A&A...460..695T}, which has a dynamical trace-back age of $18.5\pm2$~Myr \citep{2020A&A...642A.179M}.
The debris disc of AU Mic \citep{2004Sci...303.1990K} was first detected thanks to IRAS 12, 25, and 60 $\mu$m infrared observations \citep{1991A&A...244..433M}. The disc is seen edge-on and has a complex inner structure, with rich substructures at the scale of 1~AU, a central cavity in the inner 30 AU that lacks small grains, and a prominent loop that rises 2.3\,AU above the disc midplane \citep[see][for an extensive review]{2019ApJ...883L...8W}. These structures have been interpreted as possible tracers of ongoing planet formation. 

Combining light curves from NASA's Transiting Exoplanet Survey Satellite (TESS) with near-infrared radial velocity measurements, \citet{2020Natur.582..497P} recently announced the discovery of an 8.5-day transiting planet (AU Mic~b) that lies within the inner cavity of the disc. These authors also found evidence for the presence of an additional planet candidate at $\sim$30\,days. The second planet in the system (AU Mic~c) with an orbital period of about 18 days was recently confirmed based on \textit{TESS} data by \citet{2021A&A...649A.177M}.    

Being a young M-type star, AU Mic is very active. The TESS light curve from Sector 1 shows frequent flares superimposed on a 0\fm08 (peak-to-peak) photometric variability associated with the presence of active regions carried around by stellar rotation \citep{2020Natur.582..497P}. The distance to AU Mic is only $9.7141\,\pm\,0.0022$~pc \citep{2018A&A...616A...1G}, making it one of the closest planetary systems with a transiting planet known to date.

{The sky-projected spin-orbit angle of AU\,Mic\,b has been measured by several authors: 
$\lambda=-2.96_{-10.30} ^{+10.44}$ \citep{2020A&A...643A..25P};
$\lambda=0_{-15} ^{+18}$
\citep{2020A&A...641L...1M};
$\lambda=-4.7_{-6.4} ^{+6.8}$
\citep{2020ApJ...899L..13H};
$\lambda=+5_{-15} ^{+16}$
\citep{2020arXiv200613675A}.
These measurements are all consistent with an aligned orbit, suggesting that AU\,Mic\,b likely formed in the protoplanetary disc that evolved into the current debris disc,} provided that the star–disc–planet components of the system share the same angular momentum orientation.

Planets transiting magnetically active stars offer a unique opportunity to directly map the distribution of surface features, such as spots and faculae, along the transit chord \citep{2010A&A...512A..38L}. The passage of the planets in front of active regions that are spatially located along the transit chord can induce transit timing variations \citep[TTVs; see, e.g.][]{2013MNRAS.430.3032B,Mazeh2015}.
Different methods have been proposed to estimate the spot-filling factor along the transit chord and correct for unocculted star spot effects \citep{2012A&A...545A.109B,2011A&A...533A..44L,2014ApJ...796..132D,2020A&A...641A..82R}. 

The CHaracterising ExOPlanet Satellite (CHEOPS) is the first ESA  space  mission primarily dedicated to  the  study of known extrasolar planets \citep{2020ExA...tmp...53B}. Its 30 cm (effective) aperture  telescope  enables the collection of ultra-high precision  time-series photometry of  exoplanetary systems in a broad optical bandpass. CHEOPS is a pointed mission, optimised to obtain high-cadence photometric observations for a single star at a time, and started its scientific operations in March 2020 \citep{2020A&A...643A..94L}.

Here we present high-precision photometric follow-up observations of AU\,Mic~b carried out with the CHEOPS space telescope. We aim to independently confirm the recently announced transiting planet AU\,Mic\,b and to improve its transit parameters and radius. The paper is organised as follows. We describe the observations in Section 2 and present the data detrending and analysis in Sections 3 and 4. We discuss our results in Section 5 and conclude in Section 6.

\begin{table*}
    \centering
    \caption{CHEOPS observation logs. Time notation follows the ISO-8601 convention. {The File Key supports the fast identification of the observations in the CHEOPS archive.}}
    \begin{tabular}{cccccccc}
    \hline
    \hline
    \noalign{\smallskip}
    Visit & Start date & End date & File key & CHEOPS & Integ. & Co-added & Num. of     \\
     \# &  &  & & product & time (s) & exposures & frames   \\
    \noalign{\smallskip}
    \hline
    \noalign{\smallskip}
    1 & {\small 2020-07-10 09:07:24} & 2020-07-11 00:45:29 & {\small PR100010\_TG001701} & Subarray &30   &15\,s $\times$ 2 & 3597 \\
     & & & & {\it Imagettes}& 15 & --- & 7194 \\
    2 & 2020-08-21 19:07:06 & 2020-08-22 06:05:26 & {\small PR100010\_TG001702} & Subarray &30  &15\,s $\times$ 2 & 4422 \\
     & & & & {\it Imagettes}& 15 & --- & 8844 \\
    3 & 2020-09-24 16:29:54 & 2020-09-25 10:16:00 & {\small PR100010\_TG001801} & Subarray & 42 & 3\,s $\times$ 14  &  873 \\
     && & & {\it Imagettes} & 3 & ---  & 12222  \\
    \noalign{\smallskip}
  \hline
    \end{tabular}
    \label{tab:cheops_log}
\end{table*}

\begin{table*}
    \centering
    \caption{TESS and ASAS observation logs. Time notation follows the ISO-8601 convention.}
    \begin{tabular}{lcccccl}
    \hline
    \hline
    \noalign{\smallskip}
    Survey & Start date & End date & T$_\mathrm{exp}$ (s) & Num. points & Remarks \\
    \noalign{\smallskip}
    \hline
    \noalign{\smallskip}
    TESS S1 & 2018-07-25 19:12:03 & 2018-08-22 16:01:18 & 120  & 20066 & \cite{2020Natur.582..497P}, 2 transits  \\
    TESS S27 & 2020-07-05 18:48:10 & 2020-07-30 03:33:42 &20  & 97519 &  3 transits \\
    \noalign{\smallskip}
    \hline
    \noalign{\smallskip}
    ASAS & 2001-08-21 03:09:30 & 2008-04-18 09:34:39 & 180  & 430 &  Stellar rotation\\ 
    \noalign{\smallskip}
    \hline
    \end{tabular}
    \label{tab:log_tess}
\end{table*}

\section{Space- and ground-based photometry}

In the present paper we used CHEOPS photometry (Sect.~\ref{CHEOPS_Phot}) complemented with TESS light curves (Sect.~\ref{TESS_Phot}) and archival ground-based photometry from the \textit{All-Sky Automated Survey} (\textit{ASAS}, Sect.~\ref{ASAS_Phot}). The log of the observations is listed in Tables~\ref{tab:cheops_log} and \ref{tab:log_tess}.

\subsection{CHEOPS \textit{imagette} photometry}
\label{CHEOPS_Phot}

\begin{figure}
    \centering
    \includegraphics[width=0.9\columnwidth]{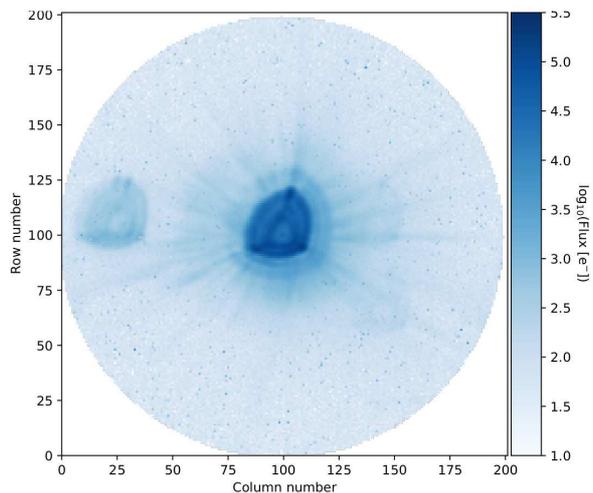}
    \caption{{\it CHEOPS} 200$\arcsec\times$\,200$\arcsec$ field of view. AU\,Mic is located at the center while the close-by faint field stars rotate during the observations.}
    \label{fig:cheops_fov}
\end{figure}

\begin{figure}
    \centering
    \includegraphics[viewport=12 44 440 400 , clip,width=8cm]{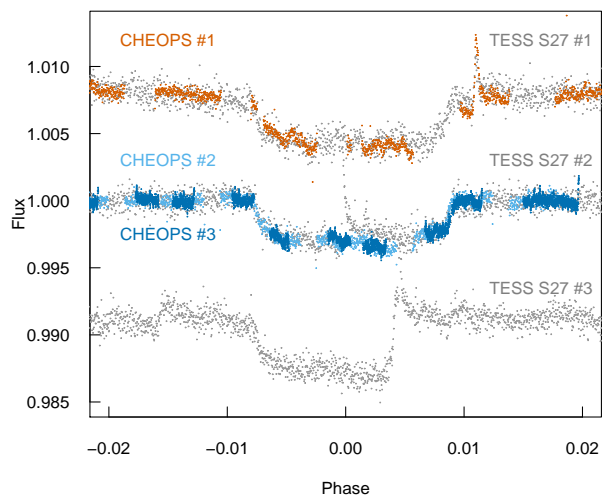}
    \caption{CHEOPS and TESS S27 observations of AU Mic in 2020. 
    CHEOPS\#{}1 and TESS S27\#{}1 were measured synchronously.
    CHEOPS\#{}2 and \#{}3 are overplotted with TESS S27\#{}2, as they were observed at the same stellar longitude at different times. TESS S27\#{}3 is plotted alone as there are no CHEOPS observations at this stellar longitude. This figure shows CHEOPS \#{}3 data points binned to 15~s cadence.}
    \label{fig:6lcs}
\end{figure}

We observed three transits of AU\,Mic\,b with CHEOPS in July, August, and September 2020. The efficiency of the observations varied between $\sim$55 and 90\,\% depending on the aspect and duration of the Earth occultation. The brightness of AU\,Mic \citep[V=8\fm6 and GAIA G=7\fm843;][]{2012AcA....62...67K,2018A&A...616A...1G} enabled short-cadence photometry. We scheduled continuous 15-second exposures for the first two visits. For the third visit we reduced the integration time to 3\,s to better sample and mask out flares. Figure~\ref{fig:cheops_fov} shows a 200$\arcsec\times$\,200$\arcsec$ CHEOPS sub-array image centred around AU\,Mic. The CHEOPS Data Reduction Pipeline \citep[DRP;][]{2020A&A...635A..24H} provides aperture photometry of these sub-array frames. Because of the satellite bandwidth limitation, individual short exposures are normally co-added before download. For observations of AU\,Mic, this resulted in a sub-array cadence of 30\,s, even if the cadence of individual photometric exposures is as short as 3\,s.

In addition to the sub-arrays, there are images of  30 pixels in radius centred around the target, the so-called \textit{imagettes}. These \textit{imagettes} can be downloaded at a higher cadence, typically without the need to co-add them. In order to take advantage of the higher temporal cadence of the \textit{imagettes} compared to the subarrays, we developed a custom \textit{PSF imagette photometric extraction} (PIPE) tool that derives a point-spread function (PSF) from the data series, and fits it to each \textit{imagette} to extract the photometry. The PIPE photometry is consistent with that derived from the DRP, with comparable performance. The higher cadence is particularly useful for resolving flares in active stars such as AU\,Mic.

In order to keep the cold plate radiators facing away from the Earth, the spacecraft rolls during its orbit, causing the field of view to rotate around the pointing direction. The target star remains stationary to within typically a pixel \citep{2021arXiv210100663B}. The irregular shape of the CHEOPS point-spread function (Fig.~\ref{fig:cheops_fov}) along with the rotation of the field of view produces a variation of the flux contamination from nearby sources in phase with the roll angle of the spacecraft \citep{2020A&A...635A..24H, 2020ExA...tmp...53B}. We accounted for the roll angle systematics affecting the CHEOPS light curve of AU Mic using a fourth-order Fourier polynomial, with coefficients in the range of 0--400 ppm, as described in the Appendix. The raw CHEOPS light curves are shown in Fig.~\ref{fig:6lcs}.

\subsection{TESS data}
\label{TESS_Phot}

TESS photometrically monitored AU\,Mic twice. Two transits were observed in Sector 1 (S1, July-August 2018) with a 120-s cadence and presented in \citet{2020Natur.582..497P}. Three additional transits were recently observed by TESS during its extended mission in Sector 27 (S27, July 2020) with a 20-s cadence. We retrieved the S1 and S27 TESS light curves from the Mikulski Archive for Space Telescopes (MAST) portal\footnote{\url{https://mast.stsci.edu/portal/Mashup/Clients/Mast/Portal.html}.}. For both sectors, we downloaded the simple aperture photometry (SAP) light curves detrended with the pre-search data conditioning (PDC) pipeline \citep{Stumpe2012}, that is, the so-called PDCSAP fluxes. 

The first transit observed by TESS in Sector 27 (hereafter transit S27\#1) was also observed by CHEOPS during its first visit (hereafter CHEOPS\#1). The different band-passes of the two instruments allowed us to study the effects of stellar activity on the estimate of the planetary radius, as described in the following sections.

\subsection{ASAS photometry}
\label{ASAS_Phot}

We also used archival ground-based photometry from the \textit{ASAS} \citep[ASAS,][]{1997AcA....47..467P}. ASAS is currently monitoring $\sim$10$^7$ stars brighter than V\,$<$\,14 using two wide-field cameras in Chile (since 1997) and in Maui (since 2006). We retrieved the ASAS time-series photometry of AU Mic from the web-page of the project\footnote{\url{http://www.astrouw.edu.pl/asas/}.}. The data set consists of 423 data points covering a time-span of about 7.5 years, {from August 2001 to April 2008}. Following visual inspection of the light curve, we removed four obvious outliers. The average scatter of the remaining 419 data points is 0.023 mag, which hides the transits but contains the photometric variability induced by the presence of stellar spots appearing and disappearing as the star rotates about its axis.

\section{Removal of systematics and flares}

\begin{figure}
    \centering
    \includegraphics[width=8cm]{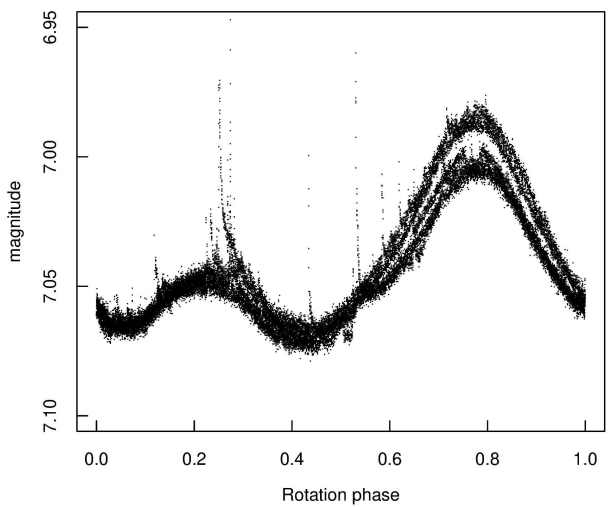}
    \caption{TESS S27 light curve of AU Mic phase-folded at the stellar rotation period ($P_{\mathrm{rot}}$=4.8367~d, $T_0=2459058.48$).}
\label{fig:rot1}
\end{figure}

The TESS and CHEOPS light curves of AU Mic show numerous flares, with occasional peak amplitudes up to 0.1~mag superimposed on a periodic spot-induced variability of about 0.08 magnitudes \citep[see Fig.~1 in][and Fig.~\ref{fig:rot1} in this work]{2020Natur.582..497P}. Flares are among the most dominant sources of uncertainty for the parameter estimates extracted from the modelling of the transit light curves, and careful treatment of the observed transits is required. Prominent flares are observed during TESS S27\#{}1 and S27\#{}2 transits (see Fig.~\ref{fig:6lcs}), with an energetic flare occurring at the mid-time of the S27\#{}2 transit, {severely biasing} the second half of the transit light curve. During the S27\#{}3 transit, a series of flares also burst, distorting the shape of the second half of the transit (Fig.~\ref{fig:6lcs}). 

Two flares are observed before and after the CHEOPS \#{}1 transit, which are the same as in the TESS S27\#{}1 transit. An energetic single flare is also observed shortly after the transit in CHEOPS\#{}1 data with a top-to-bottom amplitude of 4700 ppm, while a weaker flare is present at the ingress, more visible (1500 ppm) in CHEOPS\#{}1 data compared to TESS S27\#{}1 photometry (1200 ppm). These flares can bias the level of the reference brightness before and after the transit. Nevertheless, one rotation farther from the transit, a `quieter' light curve segment is observed out of transit. In this case, we used points from the two orbits that are one orbit apart from the transit to fit the out-of-transit level.

As AU Mic rotates about its axis, the out-of-transit level varies slowly with a peak-to-peak amplitude of 0.08 magnitudes, as a result of the rotating spot complexes on the stellar surface. This is a slow systematic component from the point of view of the transit analysis, and as a part of the pipeline, we correct for it by (1) modelling the pattern of this contamination in synthetic images and subtracting it from the observations; and (2) involving a parametric fourth-order polynomial of the roll angle to the planet fitting procedure to make sure that the possible remnants of rolling systematics are removed from the pre-cleared signal (see the amplitude of these corrections in Table~\ref{tab:roll}).

Major flares were identified as fast-increasing and exponentially decreasing light-curve features exceeding the local flux levels by at least 5$\sigma$. After removing flares and spot-induced photometric variation from the CHEOPS light curves, we inspected the residuals and found an additional variability with an amplitude of up to 20\%{} the transit depth in CHEOPS data. The pattern is observed both in and out of transit and could be due to low-energy flares and granulation. During transits, we also expected features from occulted spots in the light curve. This variability, having a characteristic timescale of the ingress and egress duration of the transit, has a strong impact on the depth and timing measurements. Taking these systematics into account, we designed a bootstrap set of transit data and re-evaluated the precision of the planet parameter estimates, as described in the Appendix.

\section{Data analysis}

\begin{figure}
    \centering
    \includegraphics[width=0.45\textwidth]{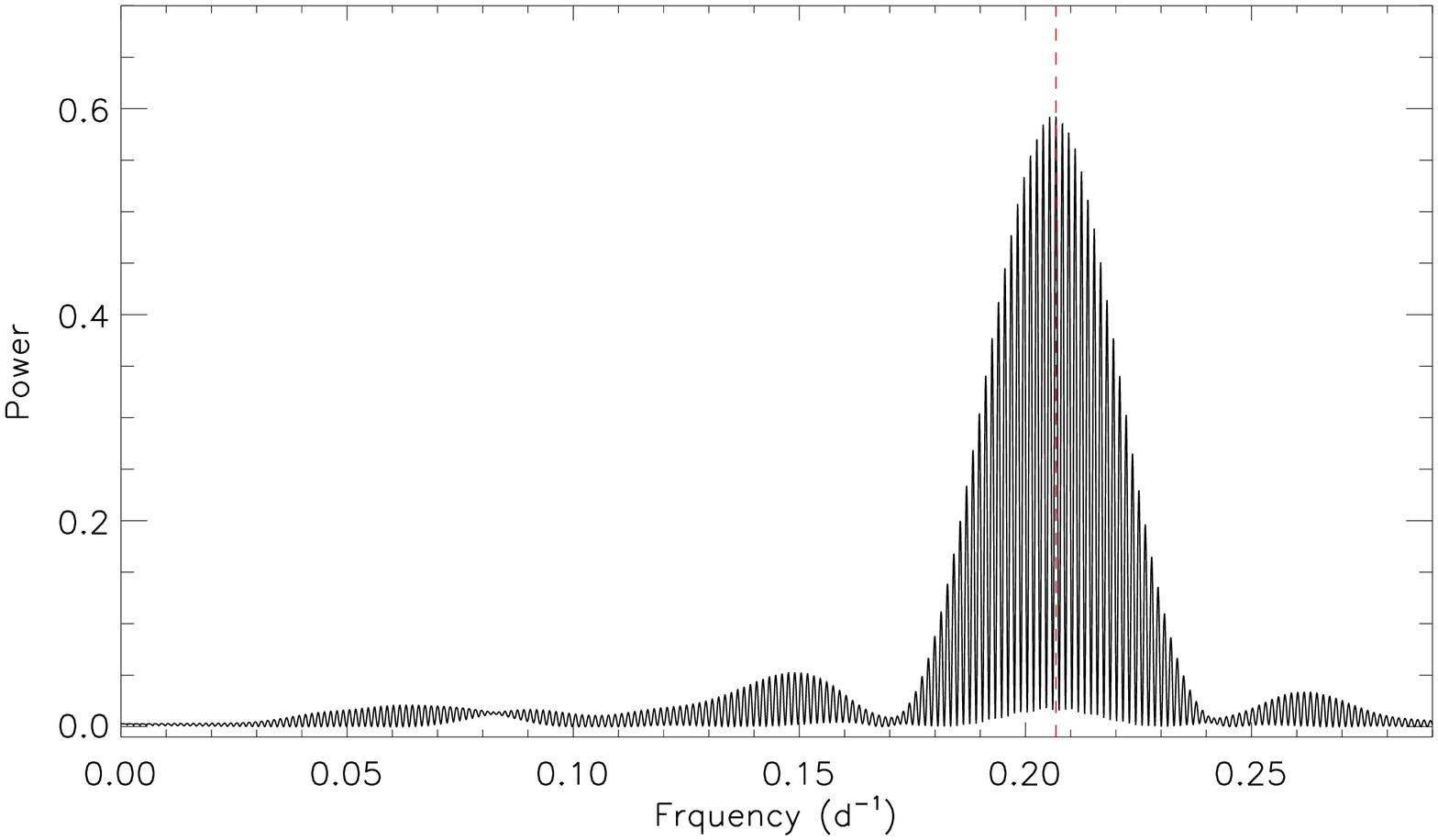}
    \includegraphics[width=0.45\textwidth]{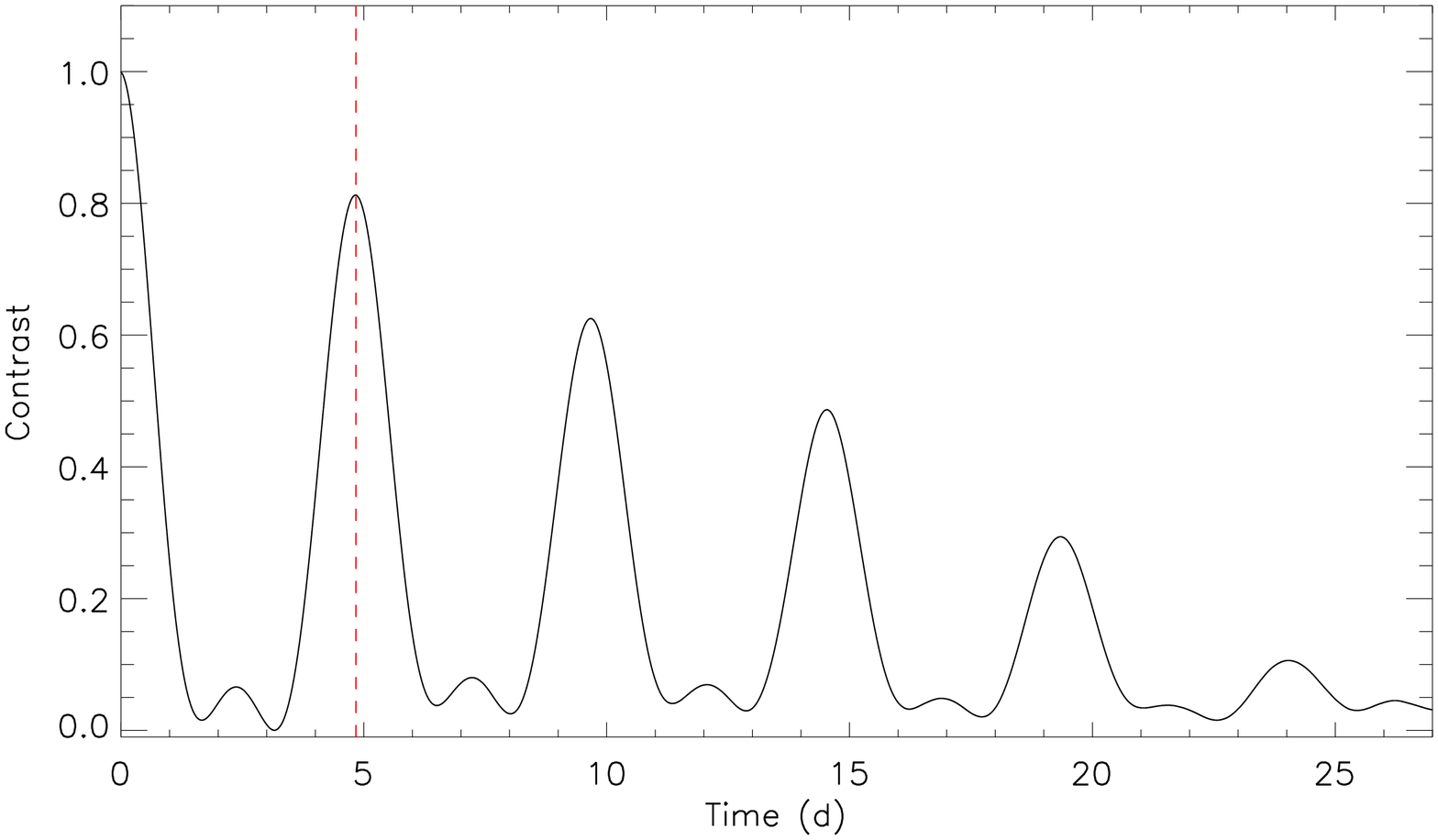}
    \caption{\emph{Upper panel}. Generalised Lomb-Scargle periodogram of the combined S1 and S27 TESS light curves. The red dashed line marks the rotation period of the star. The equally spaced peaks symmetrically distributed around the dominant frequency are due to the $\sim$2-year gap between the two TESS observations. \emph{Lower panel}: Autocorrelation function of the combined TESS light curve. The red dashed line marks the rotation period of the star. }
    \label{fig:GLS_Periodogram}
\end{figure}

\begin{figure*}
    \centering
\includegraphics[width=16cm]{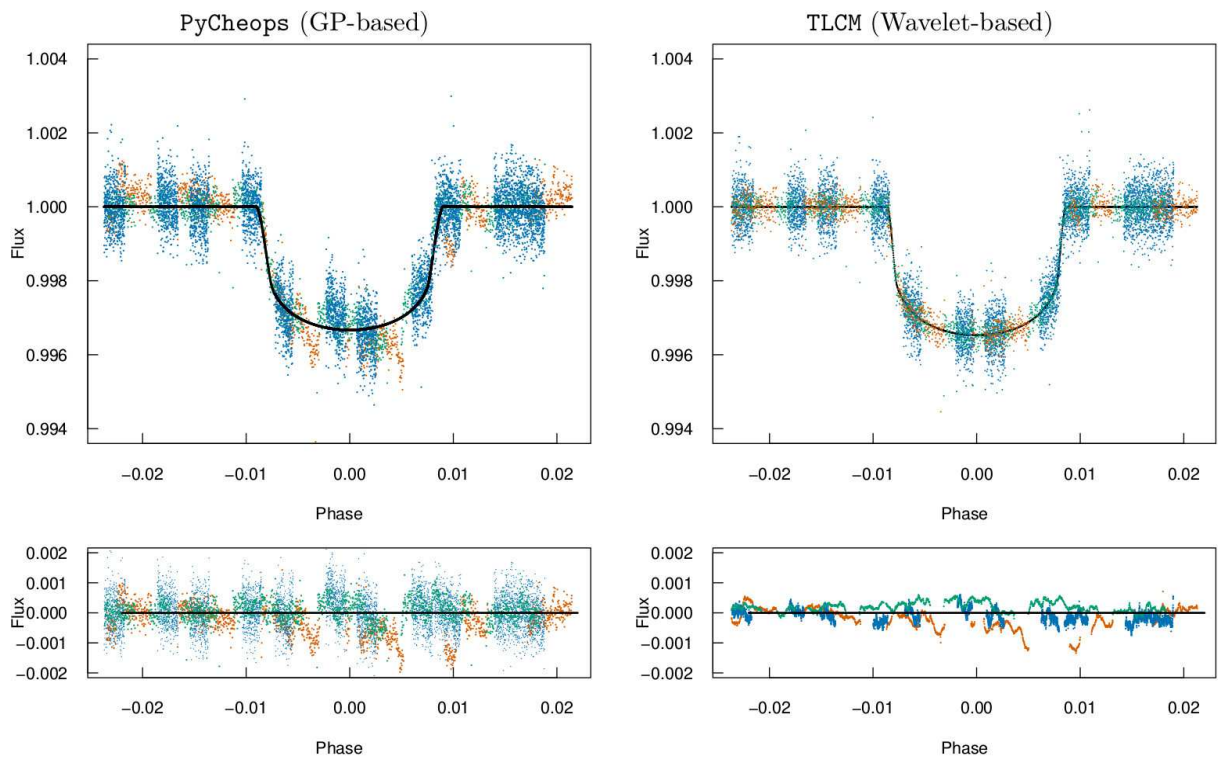}
    \caption{Joint solution of the three CHEOPS transit observations, as derived following the two methods presented in Sect.~\ref{LC_Modelling}. The colors code is the same as in Fig.~\ref{fig:6lcs}. \emph{Left upper panel}. Phase-folded CHEOPS data and best fitting transit model as obtained with \texttt{pycheops}. \emph{Left lower panel}. Residuals to the \texttt{pycheops} solution. \emph{Right upper panel}. Phase-folded CHEOPS data, \texttt{TLCM1} best-fitting transit model, and white (uncorrelated) noise component separated by the TLCM. \emph{Right lower panel}. Correlated part of the noise. The photometric variation is due to stellar activity and to the planet crossing active regions. All panels shows unbinned CHEOPS data from \emph{imagette} photometry.}
    \label{fig:fits}
\end{figure*}

\subsection{Stellar rotation}

The magnetically active star AU Mic has a projected rotational velocity of v\,sin\,i$_*$\,=\,7.5\,$\pm$\,0.9\,km\,s$^{-1}$ \citep{2020A&A...641L...1M}. A periodic pattern is observed in the TESS S1 data at a reported period of 4.86~d \citep{2020Natur.582..497P}, which is ascribable to the presence of spots combined with stellar rotation. TESS S27 data show a rotation pattern similar to S1 data from 2018 \citep{2020Natur.582..497P}, with cycle-to-cycle changes near the peak flux (Fig.~\ref{fig:rot1}). As in 2018, four stellar rotations are covered by the 2020 TESS S27 light curve.

We masked out flares and transits of AU Mic and combined the S1 and S27 TESS light curves by separately median-normalising the two time-series. The generalised Lomb-Scargle (GLS) periodogram \citep{Zechmeister09} of the combined TESS light curve (Fig.~\ref{fig:GLS_Periodogram}) displays a significant peak at 4.8362\,$\pm$\,0.0002\,d with a false-alarm probability (FAP) well below 10$^{-6}$. We estimated the FAP using the bootstrap randomisation technique \citep{Murdoch1993}. Briefly, we computed the GLS periodogram of 10$^6$ mock data sets obtained by randomly shuffling the photometric measurements and their uncertainties, while keeping the time of observations fixed. We defined the FAP as the fraction randomised data sets whose highest GLS power exceeds that of the original data set in the frequency range 0.0 -- 0.5\,d$^{-1}$. We found no false alarms out of the 10$^6$ trials, implying a FAP\,$<$\,10$^{-6}$. We note that the periodogram of the combined TESS light curves (Fig.~\ref{fig:GLS_Periodogram}) also displays a set of equally spaced aliases that are symmetrically distributed around the highest peak, with a frequency separation of $\sim$0.0014\,d$^{-1}$ ($\sim$714\,d). Those aliases are caused by the $\sim$2-year gap between the two TESS visits. We also performed a frequency analysis of each TESS time-series separately and obtained consistent results but with a larger uncertainty owing to the shorter baseline of each TESS sector (about 27 days).

We also used the autocorrelation function \citep[ACF;][]{McQuillan2013,McQuillan2014} method to measure the rotational period of AU\,Mic from the combined S1 and S27 TESS light curves. Briefly, the ACF technique measures the degree of self-similarity of the light curve over a range of lags. In the case of rotational modulation, a peak in the ACF will occur at the number of lags corresponding to the period at which the spot-crossing signature repeats. The ACF of the TESS light curve displays its first two correlation peaks at P$_\mathrm{rot}$/2$\approx$\,2.4\,d and P$_\mathrm{rot}$\,$\approx$\,4.8\,d (Fig.~\ref{fig:GLS_Periodogram}). The latter is the dominant peak at the rotational period of the star. We attributed the peak at 2.4\,d to a partial correlation between active regions at opposite longitudes on the stellar photosphere. As the evolution timescale of active regions is longer than the rotation period of the star, the ACF shows secondary correlation peaks at $3/2$\,P$_\mathrm{rot}$, $5/2$\,P$_\mathrm{rot}$, and $7/2$\,P$_\mathrm{rot}$, as well as at 2\,P$_\mathrm{rot}$, 3\,P$_\mathrm{rot}$, 4\,P$_\mathrm{rot}$, and 5\,P$_\mathrm{rot}$.  We fitted a Gaussian function to the correlation peak with the highest contrast and found a rotation period of P$_\mathrm{rot}$\,=\,4.8367\,$\pm$\,0.0006\,d.

We finally performed a frequency analysis of the ASAS light curve of AU Mic (Sect.~\ref{ASAS_Phot}). The periodogram shows its highest peak at a frequency of 0.20648\,$\pm$\,0.00013 d$^{-1}$, corresponding to a rotation period of P$_\mathrm{rot}$\,=\,4.843\,$\pm$\,0.003\,d. Although the sampling of the ASAS photometry is sparse and the noise is much higher than in case of TESS and CHEOPS observations, results are in very good agreement, suggesting that there are likely some `preferred longitudes' of starspots that give rise to a stable signal over a long time-span. Preferred longitudes have been observed to be present regularly in the case of several active stars \citep{2009A&ARv..17..251S,2007MmSAI..78..242B}, including planet-host stars
\citep{2009A&A...493..193L,2011ARep...55..341S},
and AU~Mic fits well in this picture.

The three analyses provide consistent results well within their nominal uncertainties. For further analysis, we used the results with the highest nominal precision (ACF) and adopted a rotation period of P$_\mathrm{rot}$\,=\,4.8367\,$\pm$\,0.0006\,d.\\ 

Our precise estimate of the rotational period reveals that there is a low-order commensurability between the stellar spin and the planet's orbital period in the AU\,Mic system, in the form of:
\begin{equation}
    7.0 \times P_{\rm rot} = ( \ 4.0006 \pm 0.0005 \ ) \times P_{\rm  orb}.
\end{equation}

This has important consequences for our interpretation of the measurements. As one orbit takes 1.75 stellar rotations, the stellar central longitude lags exactly $90 \pm 0.004^\circ$ in the retrograde direction from transit to transit. As a result, there are only four possible geometrical configurations for transits that are repeating continuously. Those transits that are separated by $4\times\,n\,\times\,P_{\rm rot}$ orbits are observed in front of the exact same stellar meridian, suffer very similar systematics from transited stellar features if the transit observations are taken close in time, and can be directly compared ($n$ is an integer). This is why transits taken at different times are plotted over each other in Fig.~\ref{fig:6lcs}.\\

{\cite{2021MNRAS.502..188K} used the brightness imaging technique on spectropolarimetric SPIROU data to reconstruct the differential rotation of AU Mic from spectral line profiles. They found that AU Mic has a solar-like differential rotation, with rotation periods of $\mathrm{P}_{\mathrm{eq}} = 4.84 \pm 0.01$ d at the equator and $\mathrm{P}_{\mathrm{pol}} = 5.10 \pm 0.15$ d at the poles. \cite{2021MNRAS.502..188K} also found that active regions seem to be distributed on most of the stellar photosphere (see Figure 7 in their paper), with a higher concentration around the equator.

We measured a stellar rotation period of P$_\mathrm{rot}$\,=\,4.8367\,$\pm$\,0.0006\,d, in very good agreement with the equatorial rotation period found by \citet{2021MNRAS.502..188K}. We note that our period estimate is also consistent with the expected value based on the stellar radius of R$_\star$\,$\approx$\,0.75\,$R_{\odot}$ \citep{2020Natur.582..497P}, and the projected rotational velocity of v\,sin\,i\,$\approx$\,7.5\,km\,s$^{-1}$ \citep{2020A&A...641L...1M}, assuming that the star is seen equator-on \citep{2019ApJ...883L...8W}. We are therefore confident that our estimate of the stellar rotation period refers to the stellar equator. }

\subsection{Transit light-curve modelling}
\label{LC_Modelling}

\begin{figure}
    \centering
    \includegraphics[width=8cm]{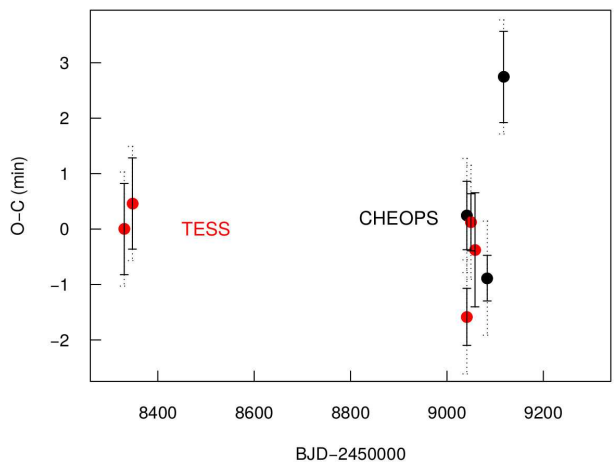}
    \caption{$O-C$ diagram of the transit mid-times derived using an orbital period of 8.462995 days {and an epoch of} $T_0=2459041.28272$~BJD. TESS and CHEOPS transits are shown with red and black circles, respectively. The solid error bars mark the fitted uncertainties (precision), while the dotted error bars show the bootstrap errors (estimated accuracy), as derived using simulated \textit{data sets}.(Sect.~\ref{Sec:ARIMA}).}
    \label{fig:oc}
\end{figure}

We determined the transit parameters using the \texttt{pycheops} software module (Maxted et al., in prep.). \texttt{pycheops} uses the \texttt{qpower2} transit model and power-2 limb-darkening law \citep[][]{Maxted2019}. It uses a Gaussian process (GP) regression to account for the roll angle systematics, whose parameters are estimated using the out-of-transit data. Flares can severely bias the roll angle models of  systematic error and also overestimate the level of correlated noise in the GP module. We therefore masked out the flares before modelling the data with \texttt{pycheops}.

{The model parameters are the transit time ($T_t$), the transit depth parameter $D = (R_p/R_\star)^2$, the transit duration parameter
$W=(R_s/a)\sqrt{(1+k)^2 - b^2}/\pi$, and the impact parameter
$b = a \cos(i)/R_s$.} 
The priors are listed in Table \ref{table:priors}. {The stellar fundamental parameters were taken from SWEET-Cat \citep{2018A&A...620A..58S}, which are the same as in \cite{2020Natur.582..497P}.} The period was kept fixed and taken from the $O-C$ analysis (see paragraph below). The noise model was calculated with \texttt{celerite} using the white-noise term $JitterTerm(\log \sigma_w)$ plus, optionally, a GP with kernel $SHOTerm(\log \omega _{0}, \log S_0, \log Q)$. The same values of $\log \sigma_w$, $\log \omega_0$, $\log S_0$ and $\log Q$ are used for all the CHEOPS data sets in the combined fit. The priors for the GP noise parameters were $S_0$ between $-30$ and $0$, $\log \omega$ between $-2.3$...$8$, and $\log\sigma$ between $-16$ and $-1$. The phase-folded CHEOPS transit light curves are shown in the left panel of Fig.~\ref{fig:fits} along with the best-fitting transit model derived with \texttt{pycheops}.

\begin{table}
    \caption{Priors used with \texttt{pycheops}, \texttt{TLCM1}, and
    \texttt{TLCM2}.}
    \centering
    \begin{tabular}{ccl}
\hline
\hline
\noalign{\smallskip}
Method & Parameter & Prior\\
\noalign{\smallskip}
\hline
\noalign{\smallskip}
\texttt{pycheops}
&$D$& N(0.003,0.001)\\
&$W$& N(0.019,0.005)\\
&$b$& N(0.01,0.17)\\
&$T_t$& N(2459041.2747,0.0022)\\
\noalign{\smallskip}
\hline
\noalign{\smallskip}
\texttt{TLCM1}
&$T_{\rm eff}$&U(3700,100)\\
&$Z$&U(0.019,0.02)\\
\noalign{\smallskip}
\hline
\noalign{\smallskip}
\texttt{TLCM2}
&$R_\star$& N(0.75, 0.03)\\
&$T_{\rm eff}$&U(3700,100)\\
&$Z$&U(0.019,0.02)\\
\noalign{\smallskip}
\hline
    \end{tabular}
    \label{table:priors}
\end{table}

We found a significant variation of the orbital period well above the nominal uncertainties, and also on the nominal epoch of $T_0=2459041.28272$~BJD, depending on which transits are actually included in the joint analysis. This is a possible sign of TTVs, and/or severe stellar systematic  errors that bias the parameter estimates, as discussed in the following sections. Therefore, we did not fit $P_{\mathrm{orb}}$ directly, but we fitted the {transit time} $T_t$ of all transits individually, extending it to TESS data from 2018 and 2020. Subsequently, we determined the period with an observed--calculated ($O-C$) analysis \citep[see, e.g.][]{2005ASPC..335....3S}. Briefly, the $O-C$ analysis compares the {observed} transit times to the {calculated} mono-periodic prediction, assuming an epoch $T_0$ and a trial period $P_t$, $C=T_0+n\cdot P_t$, where $n$ is an integer counting the number of transits from $T_0$. If the period can be refined, the point distribution will have a linear component. This way, the problem of period determination is converted to a simple linear fit. As such, all CHEOPS and TESS transit light curves are used to determine period and $T_0$. Applying the linear slope as a correction to the trial period, we obtained $P_0$, the best-fitting period of the dataset. Also, the second order (a curvature) of the $O-C$ would show a period change. Consequently, the unexplained residuals to the $O-C$ have to be read in the same way as the classical TTV diagram. Another advantage is that this method separates the determination of the period and the other planet parameters, so after this analysis $P$ can be kept fixed.

\begin{table*}
    \centering
    \caption{Best-fitting parameters of AU\,Mic\,b, as derived from the joint analysis of the three CHEOPS transit light curves. Quantities marked with an asterisk are  priors. The orbital period was derived using the three CHEOPS and five TESS transit times (see Sect.~\ref{LC_Modelling} for details). {The parameters are compared to the results of \cite{2020Natur.582..497P} (P2020) and \cite{2021A&A...649A.177M} (M2021).} }
    \label{tab:my_label}
    \begin{tabular}{|l|ll|ll|ll|r|r|}
    \hline
    \noalign{\smallskip}
    & \multicolumn{2}{|c}{\texttt{pycheops} } &  
      \multicolumn{2}{|c}{\texttt{TLCM1} } &
      \multicolumn{2}{|c|}{\texttt{TLCM2} } &  P2020 
      & M2021 
      \\
      \noalign{\smallskip}
\hline
      \noalign{\smallskip}
P [d]                   &   
  8.462995 & $\pm 3\cdot 10^{-6}$ &
  8.462995 &  (fixed) &
  8.462995 &  (fixed) &
  8.46321$\pm 4\cdot 10^{-5}$ &
  8.46300 $\pm 2\cdot 10^{-6}$\\
R$_{\rm p}$/R$_*$  &
  0.0531& $\pm 0.0023 $ &     
  0.0534& $\pm  0.0006 $  &
  0.0534& $\pm  0.0006 $  &
  0.0514$\pm$0.0013 & 
  0.0526$^{+0.0003}_{-0.0002}$\\
a/R$_*$ &  
  19.24& $\pm    0.37$ & 
  19.37 & $_{-0.18}^{+0.13}$ &
  19.10 & $_{-0.18}^{+0.13}$ &
  19.1$_{-1.6}^{+1.8}$ &
  19.1$^{+0.2}_{-0.4}$\\
W [h]  &   
  3.48& $\pm $ 0.19 & 
  3.51 & $\pm$ 0.10 & 
  3.51 & $\pm$ 0.10 & 
  3.50$_{-0.59}^{+0.63}$ &
  3.50 $\pm$ 0.08\\
R$_{\star}$ [R$_{\sun}$] &   
  0.75$^{\star}$ & $\pm$ 0.03 &        
  0.725 & $\pm $ 0.15 &
  0.75$^{\star}$ & $\pm$ 0.03 &
  0.75 (fixed) & 
  0.75 (fixed)\\
M$_{\star}$ [M$_{\sun}$] &   
  0.50$^{\star}$ & $\pm$ 0.03 &        
  0.53 & $\pm $ 0.27 &
  0.56 & $\pm $ 0.06 &
  0.50$\pm 0.03$ &
  0.50 (fixed)\\
R$_{\rm p}$ [R$_{\oplus}$] &   
  4.36& $\pm $ 0.18        & 
  4.23 &$\pm 0.8$ &
  4.38& $\pm $ 0.05       &
  4.29$\pm $0.20 &
  4.07 $\pm$ 0.17
  \\
a [AU]                       &   
  0.0678 & $\pm $ 0.0013       & 
  0.0692 & $\pm $0.0006 &
  0.0704 & $\pm $0.0006 &
  0.066$_{-0.006}^{+0.007}$ &
  0.0645 $\pm$ 0.0013\\
b                            &  
  0.09 & $\pm $  0.05       & 
  0.00& $\pm 0.13$ &
  0.00& $\pm 0.13$ &
  0.16$_{-0.11}^{+0.14}$ &
  0.18 $\pm$ 0.11\\
\noalign{\smallskip}
\hline
    \end{tabular}
\end{table*}

The three CHEOPS and five TESS transits led to the following mean period of $P_{\rm orb} = 8.462995 \pm 0.000003$~d and $T_0=2459041.28272\pm0.00033$ (Table~\ref{tab:my_label}). {We note that in comparison with the orbital period found by \citet{2021A&A...649A.177M}, this result seems to be less precise, because the error interval increased after introducing three new points to the previous distribution. This is a result of low number statistics. The errors on a quantity are themselves statistical variables too, and they suffer transient fluctuations before they start converging asymptotically. In the case of CHEOPS \#1 and CHEOPS \#2 observations, the transit times coincide well (within $\sim$1 minute) with the linear fits to the TESS points, but CHEOPS \#3 is very inaccurate (2.9 minutes from the linear fit to the other points). This measurement increased the empirical scatter and scaled up the estimates on the error of the orbital period ---in a way that is consistent with the statistical behaviour of quantities and their errors.} The $O-C$ diagram is shown in Fig.~\ref{fig:oc} and is discussed further below in detail. For the interpretation of the $O-C$ diagram, and the proper comparison of our parameter estimates with those reported in the discovery paper  \citep{2020Natur.582..497P}, a critical assessment of the effects of the systematics and stellar activity is necessary.\\

We also modelled the CHEOPS transit light curves with the wavelet-based Transit and Light Curve Modeller \texttt{(TLCM)} \citep[\texttt{TLCM};][]{csizmadia20}. \texttt{TLCM} uses the light curve model of \cite{mandelagol02} along with the quadratic limb-darkening law. Instead of directly fitting the linear ($u_a$) and the quadratic ($u_b$) limb-darkening coefficients, we fitted their combinations $u_{+} = u_a + u_b$ and $u_{-} = u_a - u_b$, as they are less degenerate with the other model parameters and with each other than $u_a$ and $u_b$ \citep{brown01}.

\texttt{TLCM} accounts for the red noise via a wavelet-based approach \citep{carterwinn09}. This confers the advantage that it requires only two additional parameters, namely $\sigma_w$ (white noise level) and $\sigma_r$ (red noise factor)\footnote{See \citet{carterwinn09} for an explanation of their meaning.}. We estimated the red noise from the out-of-transit photometry. We added a four-order truncated Fourier-series to account for the baseline-variation caused by the rotation of the field of view around the telescope line of sight \citep[for details on the baseline fitting refer to][]{2020A&A...643A..94L}, which causes the stellar images to fall on different pixels every $\sim$98.7 minutes \citep[i.e., the CHEOPS orbital period;][]{2020ExA...tmp...53B}.
The pixels can vary from visit to visit as the satellite is repointed to a new target and then back to AU Mic. We therefore used different coefficients for every CHEOPS visit.

The model parameters are the scaled semi-major axis ($a/R_\star$), the planet-to-star radius ratio ($R_{\mathrm{p}}/R_\star$), the impact parameter ($b$), epoch, the noise parameters $\sigma_w$ and $\sigma_r$, and a zero-point flux level correction $p_0$. In addition, we fitted four cosine and four sine terms ($3$ transits $\times$ $8$ free parameters) to model the baseline variation. \texttt{TLCM} also determines a stellar model based on $T_{\rm eff}$ and the fitted stellar density $\rho_\star = 3 \pi (a/R_\star)^3 G^{-2} P_{\rm orb}^{-2} (1+q)$, where the mass ratio $q$ can be approximated with zero in the case of a planet. The $T_{\rm eff}$ --$\rho$ isochrones are then fitted from \cite{2011A&A...533A.109T}, which finally gives the mass and radius of the star as a fitted parameter. 

\texttt{TLCM} first searches for the initial best-fit values using a genetic algorithm (Harmony Search) and refines the parameter estimates by simulated annealing. The refined values are used to initialise the Markov chain Monte Carlo (MCMC) analysis. For the analysis of the CHEOPS transit light curves of AU Mic, we ran 500 MCMC chains with a thin factor of 100, until convergence was reached (this required 400 000 steps in each chain). The convergence was checked via monitoring the Gelman-Rubin statistic of the chains and via the estimated sample size. Table~\ref{tab:my_label} lists the inferred transit parameters and their uncertainties, which are defined as the median and 68\,\% region of the credible interval of the posterior distributions for each fitted parameter. The nominal errors are lower in the case of \texttt{TLCM models} because of the separation of white and red noise, but the accuracy of the two models is supposed to be identical.

A strong flare occurred after CHEOPS transit \#{}1, which significantly increased the out-of-transit flux after the event. 
In principle, the red noise treatment implemented in \texttt{TLCM} is able to correct for this flux change but not perfectly. To assess the impact of the flare on the planet-to-star radius, we modelled the transit twice, first including and then masking out the data points affected by the flare. We find that the planetary radius decreases by 1.8\,\% if the data points affected by the flare are removed from the light curve. As the 1$\sigma$ relative uncertainty of the planetary radius is about 1.05\,\%, this flare introduces less than 2$\sigma$ systematic uncertainty into the planet-to-star radius ratio. As for the remaining transit parameter, the differences are not significant, being less than 2$\sigma$. To account for the possible bias introduced by flares, we increased the $\sim$1\,\% relative uncertainty of the planet-to-star radius ratio to $2$\,\%.\\

The parameter estimates obtained with \texttt{pycheops} and \texttt{TLCM1} are consistent within their 1$\sigma$ uncertainties, {and they are also consistent with those of \cite{2020Natur.582..497P} and \cite{2021A&A...649A.177M}. Unbiasing the best-fitting planetary radius for systematic errors is a non-trivial task, but the compatibility of the results in this paper and the previous literature confirms both our solutions and the parameters in the previous papers.} While the stellar mass and radius estimated by \texttt{TLCM} from the isochrones and the transit light curve are in good agreement with those given by \citet{2019AAS...23325941W} and \citet{2020Natur.582..497P}, the former are less precise than the latter, leading to a larger uncertainty on the planet's radius derived by \texttt{TLCM}.
This is due to the fact that PMS stellar isochrones of M-type low-mass stars are very close to each other.

To account for the lack of precision on the stellar mass and radius inferred using the isochrones, we performed a second analysis with \texttt{TLCM} (see Table~\ref{tab:my_label}) using a Gaussian prior for $R_\star$ of N(0.75, 0.03). The stellar mass $M_\star$ was derived via the $\log \rho_\star$--$T_{\rm eff}$--metallicity relationship based on the PMS isochrones of \cite{2011A&A...533A.109T}. The results are reported in Table~\ref{tab:my_label}.

The parameter estimates derived using the three methods described in the previous paragraphs agree well within their nominal 1$\sigma$ uncertainties. We adopted the results obtained with the second \texttt{TLCM} modelling for further analyses as they have the highest precision (\texttt{TLCM2} in Table~\ref{tab:my_label}). This provided a planetary radius of R$_{\rm p}=4.38 \pm 0.05$~R$_{\oplus}$, in agreement with the value reported by \citet{2020Natur.582..497P}.

\subsection{Accuracy of the planet parameters in the presence of systematics}
\label{Sec:ARIMA}

It is known that systematics on timescales comparable to the transit duration can lead to biased parameter estimates with uncertainties that are underestimated in the formal precision produced by the analysis. To estimate a realistic accuracy in the fundamental parameters, we designed a parametric bootstrap and measured the stability of the reconstructed transit parameters.

The noise model, including stellar and instrumental systematics, was simultaneously synthesised by fitting third-order \texttt{ARIMA} models \citep{ARIMAbook,2018JPhCS1028a2189S} to the out-of-transit data after removing the low-frequency variation due to the stellar rotation {signal} and masking out evident flares as described in Sect. 2.4. \texttt{ARIMA} models produce noise terms with partial correlation, simulating systematic patterns in a parametric auto-regressive process with a moving average. The model transit light curve for AU Mic b, with the same input parameters as those obtained in Sect.~\ref{LC_Modelling}, was injected into 100 different realisations of the \texttt{ARIMA} noise model. Earth occultations were taken into account by omitting short segments of the light curve that are one CHEOPS orbit apart, with an efficiency of 85\%{}. 

The transit parameters were then determined from 100 such simulations, with each fit being solved for transit depth, duration, and transit time. The resulting parameters were compared with the input transit parameters. The transit time in the model fits has a standard deviation of 3 minutes, while the relative standard deviations were 5.5\,\%{} in the nominal value of the relative radius and 2.0\,\%{} or 4.2 minutes in transit duration, respectively. 

We conclude that the transit time can be determined from observations with an accuracy of  3 minutes at best. This means that fitted transit times can be expected to vary on a timescale of $\pm$ 3 minutes simply due to the presence of correlated noise, without any specific physical variations in the AU Mic system.

\subsection{Transit depth variation}

\begin{table}
    \centering
        \caption{Transit times and planet-to-star radius ratios derived with \texttt{TLCM1} from the individual CHEOPS transits.}
    \begin{tabular}{|l|l|l|}
    \hline
    \noalign{\smallskip}
Transit      &  $T_t$ (BJD) &  $R_p/R_\star$ [\texttt{TLCM}] \\
\noalign{\smallskip}
\hline
\noalign{\smallskip}
CHEOPS \#{}1 & 2459041.2828$\pm 0.0006$ &  0.0575$\pm 0.0012$ \\
CHEOPS \#{}2 & 2459083.5970$\pm 0.0004$ &  0.0532$\pm 0.0010$ \\
CHEOPS \#{}3 & 2459117.4515$\pm 0.0008$ &  0.0525$\pm 0.0012$ \\
\noalign{\smallskip}
\hline
    \end{tabular}
    \label{tab:transits_individual}
\end{table}

\begin{figure}
    \centering
    \includegraphics[bb=10 100 440 370, width=0.48\textwidth]{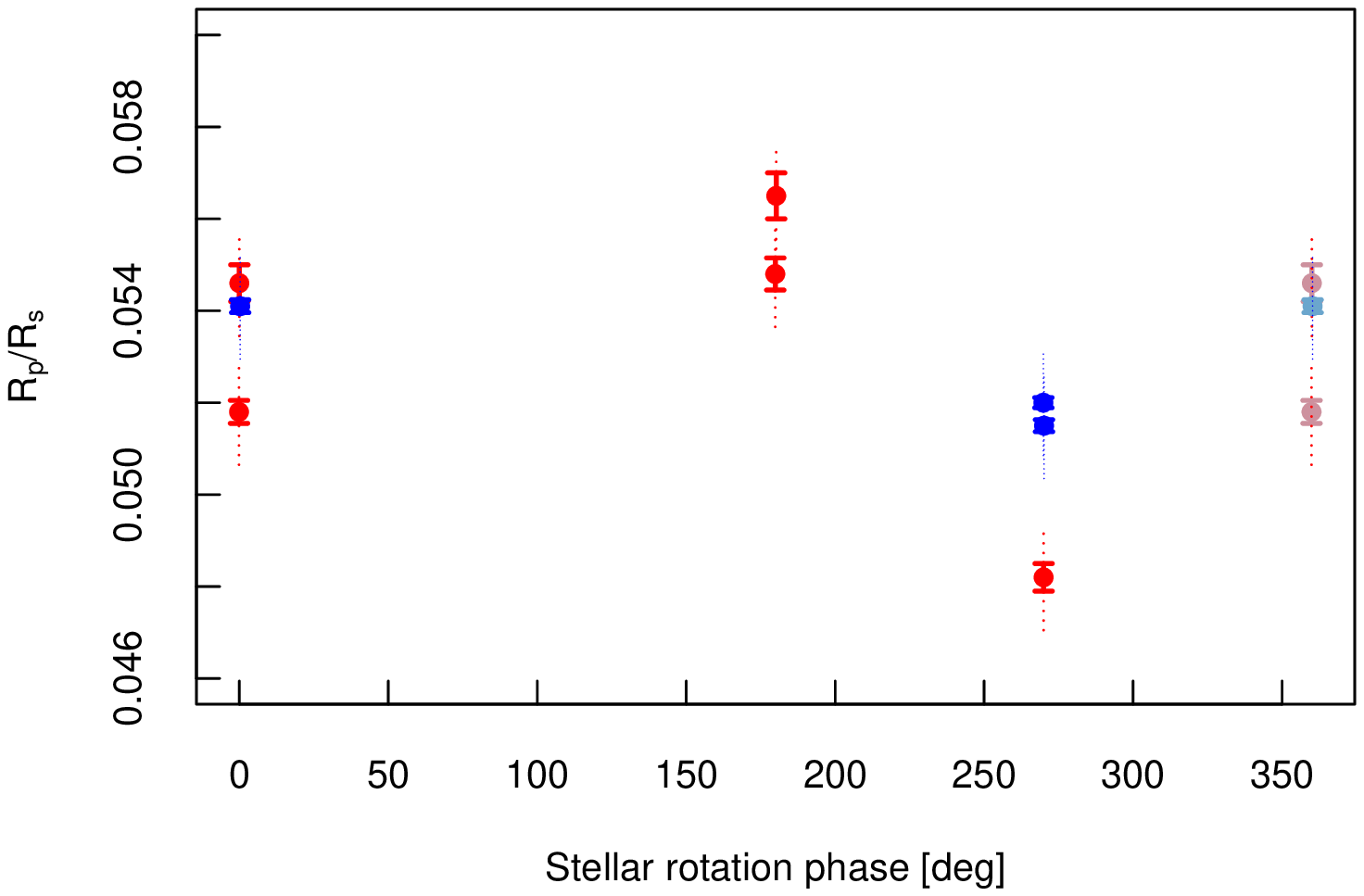}\\
    ~~\includegraphics[ width=0.463\textwidth]{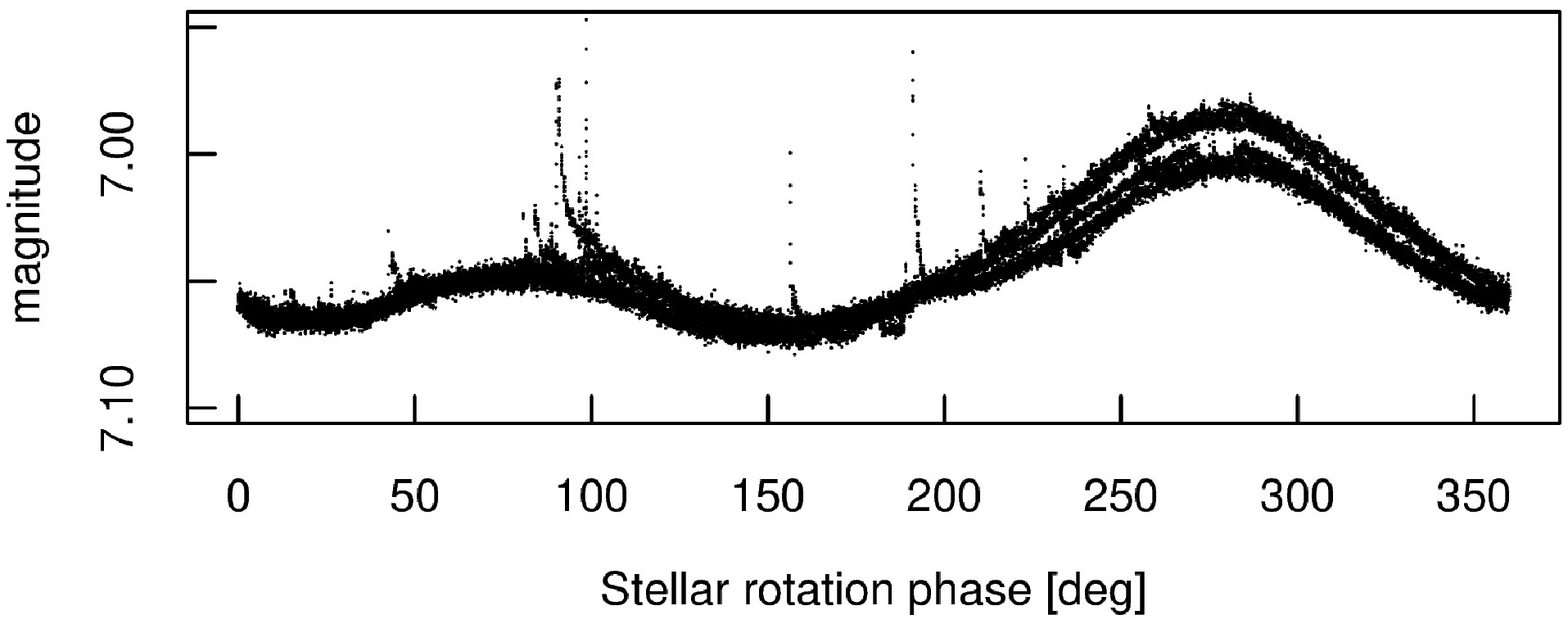}
    \caption{Planet-to-star radius ratio versus stellar rotation phase. The red and blue circles mark the TESS and CHEOPS transits, respectively, as modelled with \texttt{pyaneti}. {(We note that the data points at phase 0$^\circ$ are repeated at phase 360$^\circ$ for clarity.)} The solid error bars are the nominal uncertainty (precision), while the dotted lines are the uncertainties after accounting for the lack of accuracy with the bootstrap method described in Sect.~\ref{Sec:ARIMA}.}
    \label{fig:RpRs}
\end{figure}

The presence of spots on the photosphere of AU Mic has a twofold effect on the transit depth, and therefore on the measurement of the planet-to-star radius retrieved from the modelling of the transit light curves. 
\begin{enumerate}
    \item Spots along the transit chord that are partially or completely occulted by the planet distort the shape of the transit light curve: While the planet is passing in front of a spot, the fractional loss of light is temporarily reduced and a positive `bump' is observed in the transit light curve \citep[see, e.g.][]{Sanchis-Ojeda2011b}. 
    \item Spots that are not spatially located along the transit chord by the time the planet passes in front of the star reduce the brightness of the unocculted part of the stellar hemisphere with respect to a spot-free surface, making the transit appear deeper \citep[see, e.g.][]{Pont2008}. As the transit depth is proportional to the filling factor of the unocculted spots, transits are expected to be deeper when the star is closer to its photometric minimum.
\end{enumerate}
Both effects depend on the spot temperature contrast with respect to the temperature of the quiet photosphere, with their magnitude increasing at shorter wavelengths.
We measured the depth and the mid-time for each transit observed by CHEOPS and TESS using the code \citep[\texttt{pyaneti};][]{Barragan2019}. We fixed the period, scaled semi-major axis, impact parameter, and limb-darkening coefficients to the values derived from the joint analysis, while allowing for the mid-time and depth to vary (results are listed in Table~\ref{table:pyaneti}). Figure~\ref{fig:RpRs} displays the planet-to-star radius ratio $R_\mathrm{p}/R_\star$ as a function of the rotation phase of the star. The thick lines mark the nominal uncertainties (precision), while the dotted lines are the uncertainties as derived using the bootstrap method described in Sect.~\ref{Sec:ARIMA}. We arbitrarily set the stellar rotation phase to zero at the mid-time of the first transit observed by TESS, which occurred around one of the two photometric minima of the star. Given the 7:4 commensurability between the rotation period of the star and the orbital period of AU Mic\,b, the transits can only be observed at four rotation phases. 

We find significant transit depth variations of up to $\sim$20\,\%, well above the nominal uncertainties (solid error bars). The deepest transits are observed around the photometric minima of the star (phase 0.75 in Fig.~\ref{fig:RpRs}), when the spot filling factor is higher. We also detect a possible chromatic effect. The transits observed by CHEOPS tend to be deeper than those observed by TESS, in agreement with the hypothesis that if the star is heavily spotted, the transit appears deeper in the CHEOPS passband, which is bluer than that of TESS (Fig.~\ref{fig:RpRs}).

\subsection{Active regions along the transit chord }

The modelling of the CHEOPS transit light curves revealed the presence of systematic patterns in the residuals (Fig.~\ref{fig:fits}, lower panels). These are partly due to low-energy flares and granulation, but they can also be associated to the occultation of starspots and faculae spatially located along the transit chord.

Dark and bright spot-crossing events have already been observed in several transiting systems, such as CoRoT-2b \citep{Valio2010}, HAT-P-11 \citep{2010AAS...21531707S}, WASP-4 \citep{Sanchis-Ojeda2011a}, Qatar-2 \citep{2017MNRAS.471..394M,Dai2019}, and WASP-19 \citep{2019MNRAS.482.2065E}. These anomalies have been used to measure the spin-orbit {misalignment}, that is, the angle between the stellar rotation axis and the angular momentum vector of the planet. For instance, the observed patterns of these events in the light curves of CoRoT-2 \citep{Nutzman2011}, Kepler-17 \citep{2011ApJS..197...14D}, and WASP-85 \citep{2016AJ....151..150M}  revealed a spin-orbit alignment and a prograde orbital motion of their planets, while HAT-P-11b \citep{Sanchis-Ojeda2011b} and WASP-107b \citep{2017AJ....153..205D} have been found to have  orbital planes that are misaligned with respect to the orbital axes of their host stars. 

The AU Mic system is unique in that not only are the observed light curve features clearly separated, but the entire transit floor is continuously undulated. Whether these structures are caused by dark spots or correspond to bright faculae is unclear. As we always see the presence of active areas in the transit chord, the precise radius determination of AU Mic can only be expected from a parallel photometric (preferably both visual and infrared) and spectroscopic measurement where the map of the surface can be modelled with Doppler tomography. This ambiguity explains the larger error in $R_p/R_\star$ from the present analysis (Table~\ref{tab:my_label}) than that in \cite{2020Natur.582..497P}.

\subsection{Transit-timing variation revealed by occulted active regions}
\label{TTV_revealed}

\begin{figure}
    \centering
    \includegraphics[viewport=15 135 438 350,clip,width=8cm]{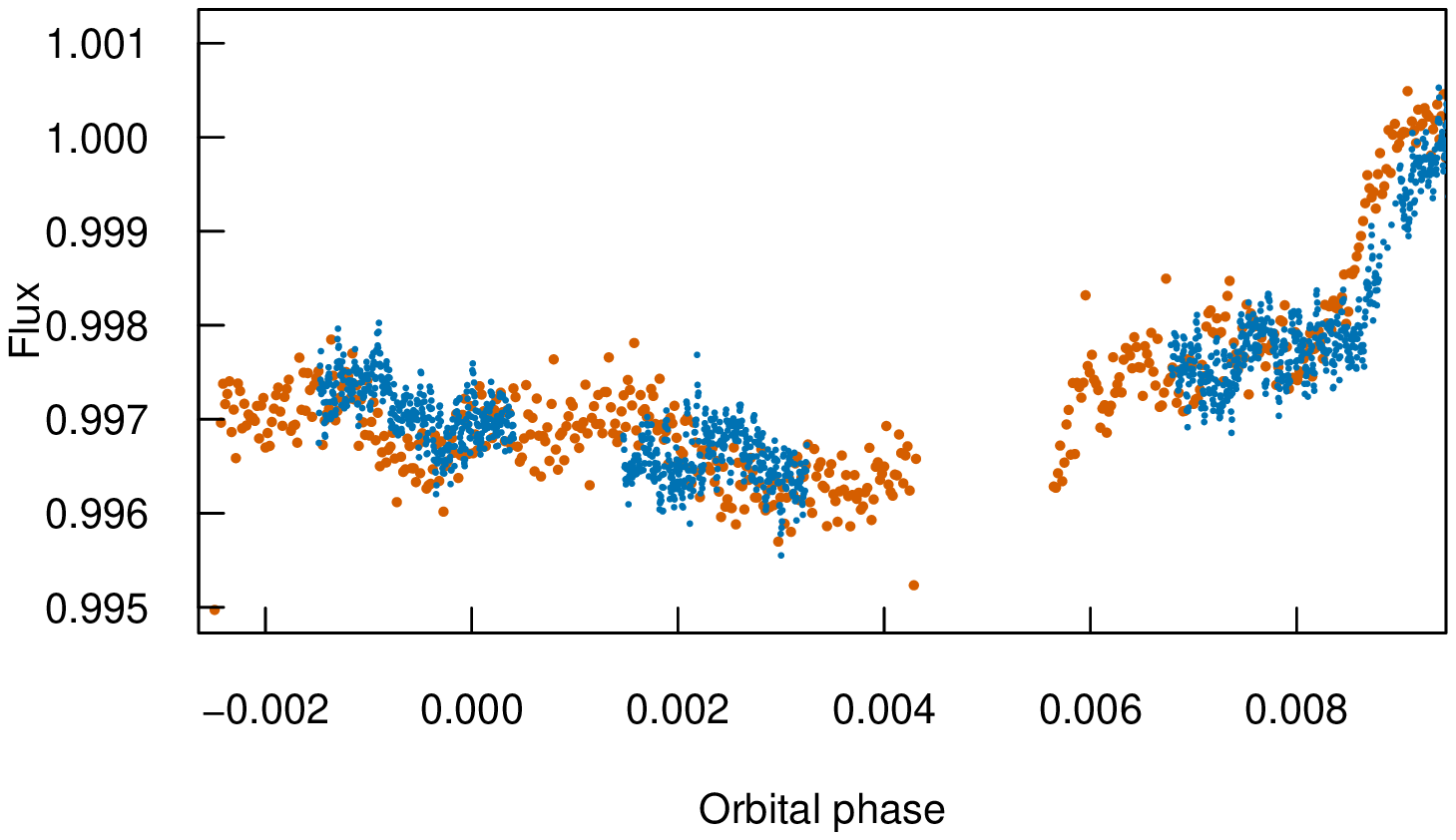}
    \includegraphics[viewport=15 100 438 350,clip,width=8cm]{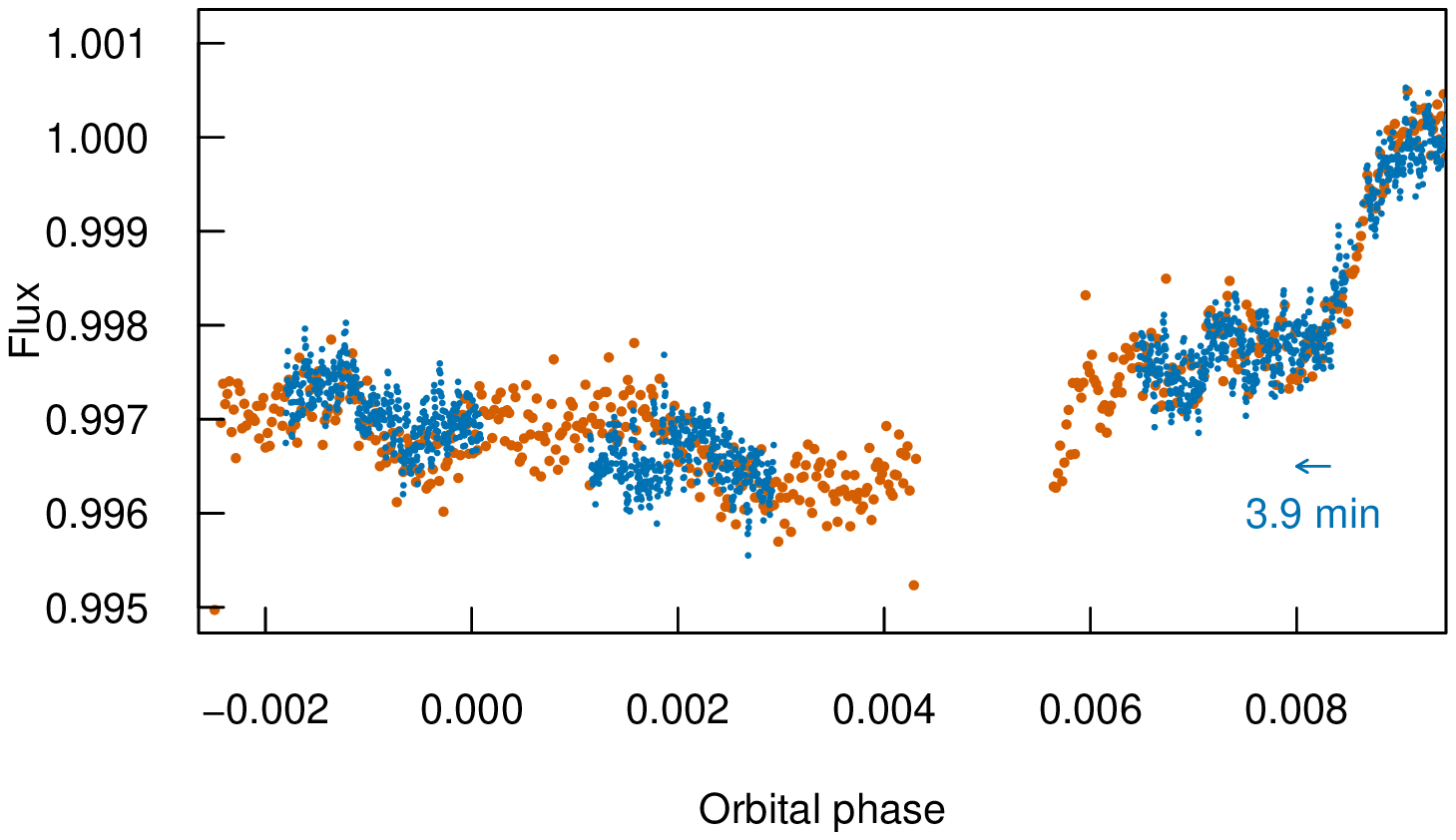}
    \includegraphics[viewport=15 136 438 314,clip,width=8cm]{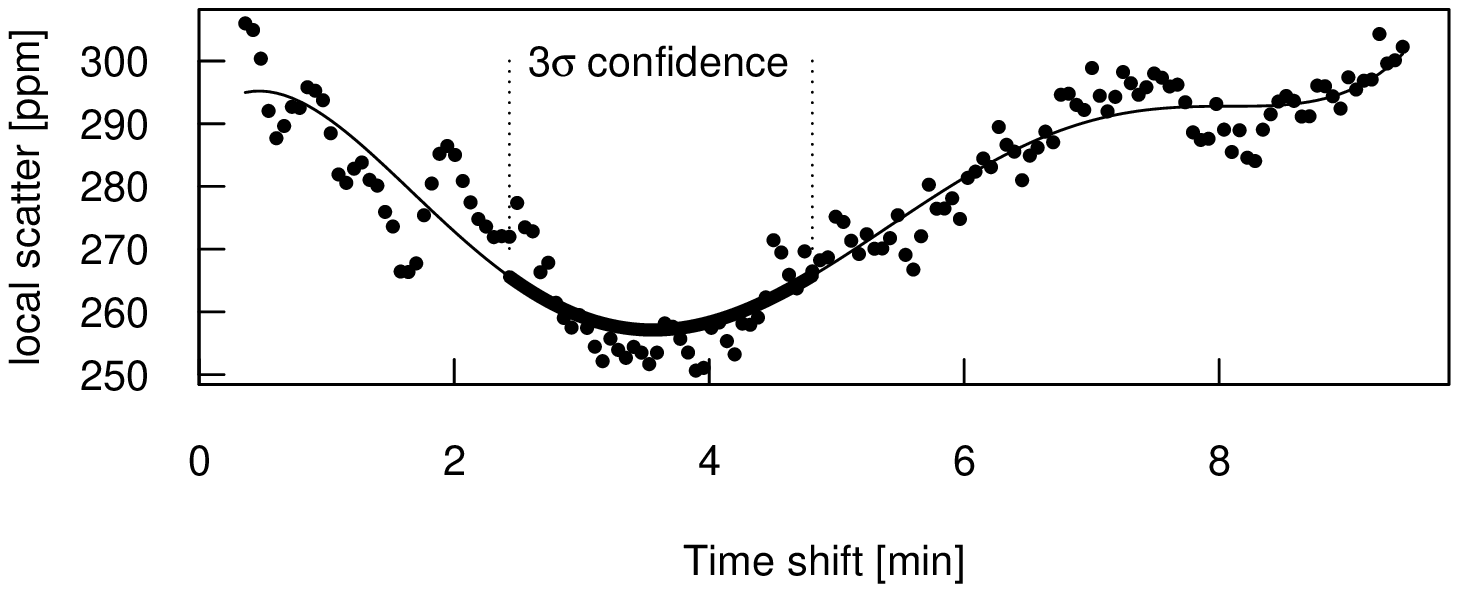}
    \caption{\emph{Top panel}:  Zoom into the second half of CHEOPS\#{}2 and CHEOPS\#{}3 transits (orange and blue data points, respectively) plotted against each other, according to the phases derived from the best-fitting joint solution (Sect.~\ref{LC_Modelling}). The colour code is the same as in Fig.~\ref{fig:6lcs}. \emph{Middle panel}: Same as in the upper panel, but with a leftward TTV correction of 3.9 minutes applied to the CHEOPS\#{}3 transit (blue data points). We note that both the egress and the structures due to occulted active regions overlap, confirming the time shift. \emph{Bottom panel}: Determination of the time-shift with a phase dispersion minimisation analysis (see text in Sect.~\ref{TTV_revealed}).}
    \label{fig:ttv}
\end{figure}

\begin{figure}
    \centering
    \includegraphics[viewport=15 135 438 350,clip,width=8cm]{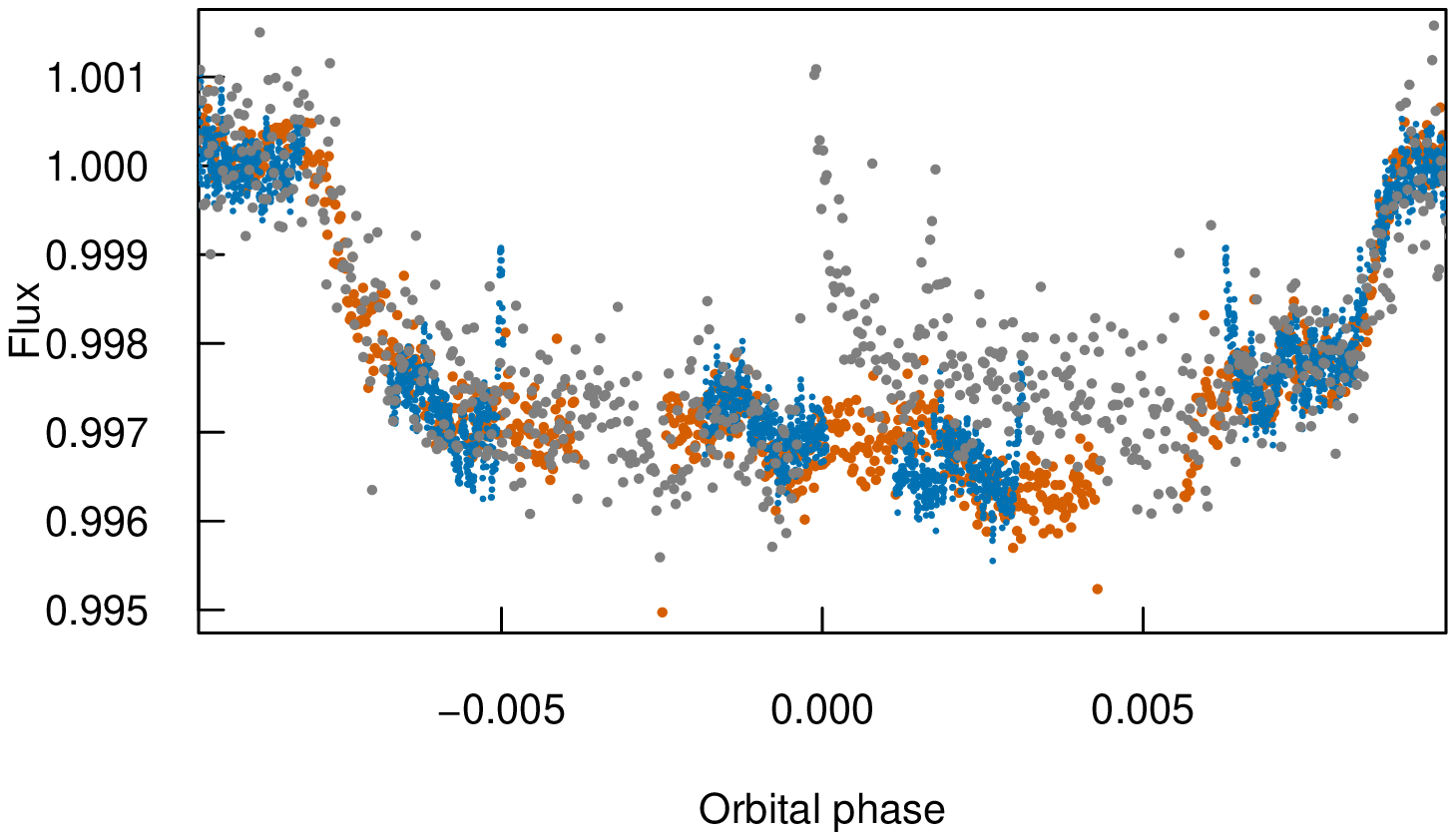}
    \includegraphics[viewport=15 136 438 314,clip,width=8cm]{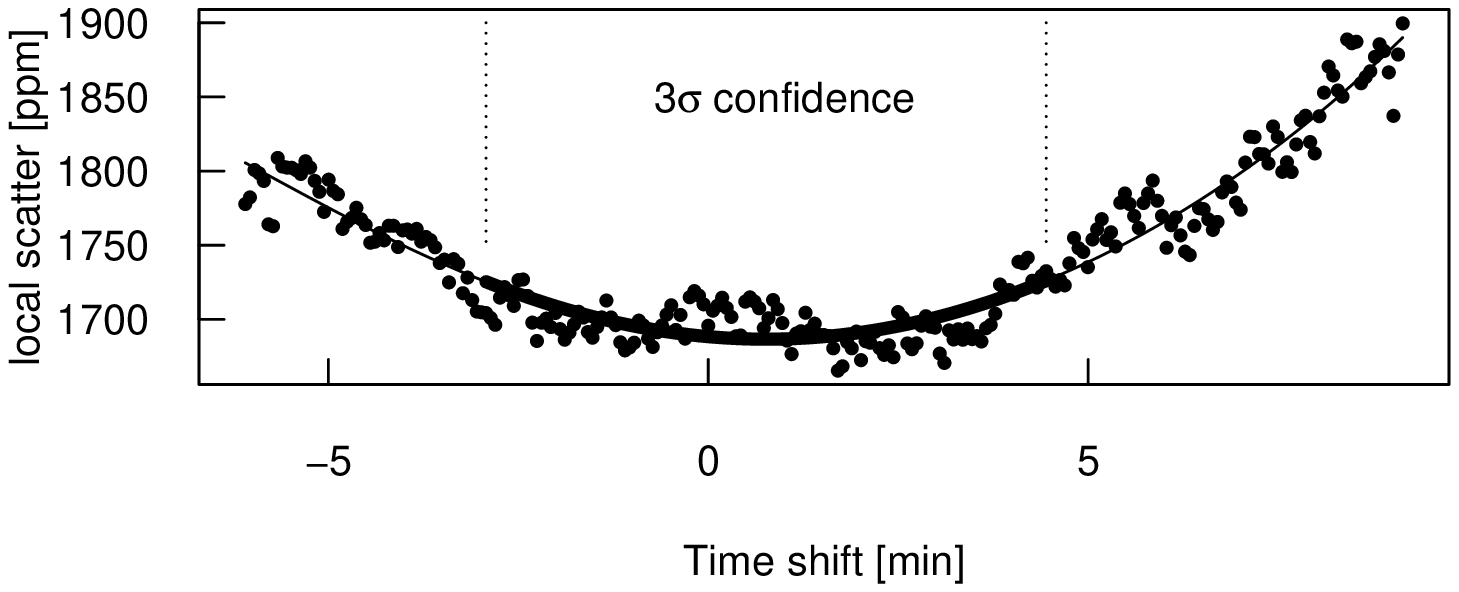}
    \caption{{\emph{Upper panel}: Unified CHEOPS \#2-\#3 transit light curves (orange and blue data points respectively). CHEOPS \#3 transit has been corrected for TTV to the time frame of CHEOPS \#2, as in the middle panel of Fig.~\ref{fig:ttv}. TESS S27 \#2 data not corrected for TTV are displayed with grey points. The colour code is the same as in Fig.~\ref{fig:6lcs}. \emph{Lower panel}: Corresponding cross-correlation analysis.}}
    \label{fig:ttv_tess}
\end{figure}

As occulted active regions  systematically distort the light curves, they are candidate sources of apparent TTVs without a dynamical background \citep{2013MNRAS.430.3032B}. An especially active system such as AU Mic has to be handled with extraordinary care, avoiding over-interpretation when comparing transit times before different stellar surface structures.

Fortunately, we know that CHEOPS\#{}2 and CHEOPS\#{}3 transits occurred at the same stellar longitude, which allows us to compare the transit light curves directly. We note that by simple visual inspection, we see that the egresses of these two transits are shifted in time (see Fig.~\ref{fig:ttv} for a zoom-in of the two CHEOPS measurements). 

To test the time-shift between these light curves, we adopted a method that is very similar to the cross-correlation. CHEOPS\#{}2 and CHEOPS\#{}3 datasets were unified by shifting the CHEOPS\#{}3 data by a shift parameter that was kept free during the analysis. The unified dataset was sorted by the orbital phase variable. The free parameter in this test is the time-lag between CHEOPS\#{}2 and CHEOPS\#{}3, in minutes. The $\sigma$ mean local scatter was determined as:
\begin{equation}
\sigma =\sqrt{\left\langle \, (f_{i+1} - f_i)^2 \right\rangle },
\end{equation}
where the mean (denoted by $ \langle \rangle $) was calculated in the transit bin only. Here, $f$ is the scaled flux of the unified (original) CHEOPS\#{}2 and (shifted) CHEOPS\#{}3 dataset, and $i$ is the rank with respect to phase.

The shape of the local scatter as a function of time-lag has then been fitted by a fifth-order polynomial (lowest panel in Fig. \ref{fig:ttv}). The confidence level of the time shift {that takes the residuals of CHEOPS\#{}2 to CHEOPS\#{}3} was determined by a threshold value of the minimum of the fitted model, plus three times the standard deviation between the model and the points. This interval (lowest panel in Fig. \ref{fig:ttv}) corresponds to shifts of between {2.1 and 5.1} minutes, which are compatible with {the TTV measured from the fitted models (Fig. \ref{fig:oc}).}

The middle panel of Fig. \ref{fig:ttv} shows the phased transit light curves (second half of the transit floor and egress part magnified) after applying the time-lag correction of 3.9 minutes. We see that, not only do the egresses overlap, but the undulation from occulted stellar surface structures also perfectly match. This confirms that (1) the observed time-lag is really due to TTVs of the planet, and (2) the undulating features are due to the star. 

{We repeated the same analysis to compare the combined CHEOPS \#2 (not TTV corrected) and CHEOPS \#3 (TTV corrected to the phase frame of CHEOPS \#2) datasets versus the TESS S27 \#2 transit –- which shows a transit at the same equatorial latitude but at a different time. In other words, the combined datasets in Fig. 8 (middle panel) and Fig. 2 (middle panel, grey dots) were compared. The result of this comparison is shown in Fig 9. We see that the minimum of the correlation function is at $\approx$1 minute, showing that CHEOPS \#2 should be shifted by $1$ minute to get the best overlap of residuals with TESS S27 \#2. However, the uncertainty is 7 minutes, which is on the order of the standard deviation of the TTV of AU Mic; and therefore this comparison is not conclusive. TESS has a different cadence, different passband (less sensitive to spots), and a different precision, and is less suited to studying the effect of occulted spots in this case.}

We stress that there is also considerable out-of-transit red noise, which is characteristic of stellar activity, and has an amplitude comparable to the red noise component which is specific to the transits. Therefore, the separation of red noise from flicker and from occulted surface features is impossible based on monochromatic transit observations only, and will be investigated in a future study involving ground-based follow up observations. Although the stellar spot map of AU Mic is unknown, the reoccurring stellar surface together with the repeating light-curve features nevertheless confirm that the observed TTVs have a dynamical origin.

\section{Discussion}

\subsection{A 7:4 period commensurability between the stellar spin and the planet orbit}

Low-order spin-orbit commensurability is not unexpected in exoplanets, but there are few systems known to exhibit this feature. A reason for the low number of detections could be the difficulty in measuring the precise rotational periods from long and continuous photometry. Typically, exoplanets in the original Kepler field are the best-suited systems for this kind of study. Presently, there is no universal theoretical framework to explain any example of a low-order spin-orbit commensurability.

Regardless of the difficulties, similar systems have been known for almost a decade. \cite{2013MNRAS.436.1883W} determined the rotational period of a large sample of KOI stars, and noted an overdensity at the 1:1 resonance in the $P_{\rm rot}$-- $P_{\rm  orb}$ parameter space, and possibly another one at 2:1.
They remarked that the distribution is valid only for large planets; while there does not appear to be a relationship between stellar rotation and planetary orbital periods for planetary candidates smaller than $R_p=6\ \rm{R_\oplus}$, candidates with $R_p>6\ \rm{R_\oplus}$ tend to appear closer to a low-order commensurability.

Concerning the 1:1 commensurability, the interaction of the planet's magnetic field was suggested to enhance the activity at the stellar surface toward the planet \citep{2012A&A...544A..23L}. These latter authors also debated whether the motion of the spots truly reflects the stellar rotation in this case, or rather that they move on the stellar surface by following the planet. The example of Kepler-17 revealed a system that is close to a perfect 8:1 spin--orbit commensurability \citep{2011ApJS..197...14D}. Another intriguing case is Kepler-13, which was found to be in a 3:5 commensurability \citep{2012MNRAS.421L.122S}. In this system, the suspected reason was suggested to be a spin--orbit misalignment; at high latitudes of the orbit, the star and planet move together in a quasi-synchronous state for a considerable amount of time and, twice per rotation, the planet resides for 8 hours within 1 degree of one of the three distinguished longitudes. 
In the case of AU Mic, the explanation cannot be a spin--orbit misalignment, as we know that AU Mic b is aligned.

\begin{figure}
    \centering
    \includegraphics[viewport=18 75 436 382, width=8cm]{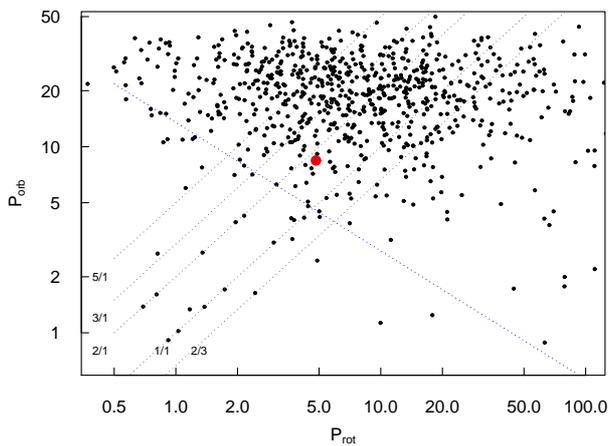}
    \caption{Position of AU~Mic (red dot) in the $P_{\mathrm{rot}}$ -- $P_{\mathrm{orb}}$ parameter space of Kepler exoplanets (small grey dots). The low-order spin--orbit commensurability ratios are shown by the black dotted lines. The lower envelope of the distribution \citep{McQuillan2013} is plotted by the blue line.}
    \label{fig:rotcomp}
\end{figure}

Another interesting feature is the dearth of close-in planets around rapid rotators, as recognised by \cite{McQuillan2013}. These latter authors also determined the empirical boundary of the lower envelope of the allowed distribution. This is a purely empirical power law that confines the majority of exoplanets, with very few outliers being observed below it. In Fig.~\ref{fig:rotcomp} we plot the sample of Kepler planets and the boundary according to \citet{McQuillan2013}, and show the position of AU Mic b in this distribution. AU Mic b is near the edge of the densely populated area, but well within the identified boundary. Interestingly, AU Mic b is a small planet, smaller than the $6\,R_\oplus$ limit \citep{2013MNRAS.436.1883W}, but the star is also smaller than most stars in the Kepler sample. Furthermore, there is an overdensity of planets in the lower left corner of Fig.~\ref{fig:rotcomp} at low-order spin--orbit resonances. For clarity we show only some of them, noting that several other possible examples can be seen by eye. 

The structure of exoplanets in the  $P_{\rm rot}$-- $P_{\rm  orb}$ plane is completely unexplained. Here, AU Mic b occupies a key location that can help us to gain a better understanding of planet--star interactions in general. 
The extreme youth of AU Mic points to a speculative interpretation that the commensurability in this system was needed for the formation of the planet.
Without a detailed theoretical explanation, we speculate that planet--star interactions may be more frequent than often recognised.

\subsection{Transit timing variations explained by perturbations from the outer planet AU Mic c}

Transit timing variations are commonly observed in multi-planet systems. The amplitude and period of TTVs is a complex question and a full solution can only be given if the periods of both the perturbing and perturbed planets are known. 
{In this section we show that the measured TTV amplitude of AU Mic b is compatible with the perturbations coming from the outer planet AU Mic c. Also, a rough estimate of its mass can be given from a TTV analysis.

Because of the scarce sampling of AU Mic transits, we have no exact information about the periodicity nor the full amplitude of the TTV. We can obtain a lower estimate of the fastest variation of TTV from the measured data. Certainly,
\begin{equation}
     \max \Big( \frac{d TTV} {dt} \Big) = 2\pi \times \frac {A(TTV)}{P(TTV)},
\end{equation}
where $A(TTV)$ and $P(TTV)$ are the amplitude and the period of the TTV, and the formula makes use of simple basic properties of the sinusoid function.

The TTV signal is thoroughly determined by \cite{2012ApJ...761..122L}; the general formulas for period and amplitude are provided in their Eqs.~7--8. The most important parameters are $j,$ which measures the closest mean-motion resonance (in the form of $j:(j-1)$), and $\Delta$, the actual distance from the perfect resonance. Taking into account the period of AU Mic b and that of AU Mic c ($18.858991\pm0.000010$~d, \cite{2021A&A...649A.177M}), we get $j=2$ and $\Delta=0.114203,$ which is precise to the last digit.
Those together determine the superperiod, that is the period of the TTV. The ratio of the orbital period of the perturber and the superperiod is $j\times \Delta\approx 0.228$, and therefore the superperiod of a TTV signal from AU Mic c is expected to be 82.6~d. 

The TTV amplitude is scaled by the $f$ parameter which has an absolute value of around unity. In the case of $j=2$, it can be calculated as $-1.190+2.20\Delta=-0.9387$ for AU Mic. 
During the gravitational interaction, the planets perturb each other's orbit predominantly through oscillations in eccentricity. This gives a correction to $f$ which is summarised in the $z_{\mathrm{free}}$ parameter. This cannot be exactly determined unless all the orbital parameters of both planets are known. However, for the known Kepler planets, it is found that $z_{\mathrm{free}}\ll\Delta$, and $z_{\mathrm{free}}$  usually means only a correction on the order of $10^{-4}$ to $f$. Assuming the same for AU Mic, the $z_{\mathrm{free}}$ correction can be neglected.

Substituting the exact value of $f$ and $j=2$ into Eq. 7 of \cite{2012ApJ...761..122L}, and converting both sides to the form of Eq. 3 in the present paper, we get
\begin{equation}
2 \pi \times \frac{A(TTV)} {P(TTV)} = 
2 \pi \frac{A(TTV) j \Delta}{P_c} = 
2.36 \frac{P_b}{P_c}\mu_c,
\end{equation}
where $P$ and $\mu$ stand for orbital period and the planet--star mass ratio, and the $\rm b$ and $\rm c$ indices indicate the planets AU Mic b and c.

The left term of Eq. 4 is $\gtrapprox7.8\times 10^{-5}$. Substituting the observed periods of AU Mic b and c into the right-hand side and the mass of AU Mic into the expression of $\mu$, we finally get $M_{\rm c}\approx 12$~$M_\oplus$. This value is only a rough estimate because we do not have information about the biases from the sparse sampling, but this result is numerically very compatible with the mass estimate for AU Mic c from its transit light curve, $2.2 M_\oplus <M_c<25.0 M_\oplus$ \citep{2021A&A...649A.177M}. This is evidence that the dominant factor in the TTV of AU Mic b is the perturbations from AU Mic c.
}

\section{Summary}

We present three transit observations of AU Mic\,b carried out with the CHEOPS space telescope, combined with the TESS light curves from Sectors 1 and 27. We analysed the transits with two different methods, which include a GP regression (\texttt{pycheops}) and a wavelet-based approach (\texttt{TLCM}). We summarise our results below.

\begin{itemize}

\item By combining the CHEOPS and TESS transit light curves, we measured an orbital period for AU Mic\,b of $P_\mathrm{orb}$\,=\,$8.462995\pm 0.000003$~d, {which confirms the previous estimate presented by \citet{2021A&A...649A.177M}.} The modelling of the three CHEOPS transits using \texttt{pycheops} and \texttt{TLCM} provides a planetary radius of $R_{\rm {p,pycheops}}\,=4.36 \pm 0.18$\,$R_\oplus$ and $R_{\rm {p,TLCM}}\,=\,4.38 \pm 0.05$\,$R_\oplus$, respectively. {These results agree within $3\sigma$ with the values found by \citet{2020Natur.582..497P} and by \citet{2021A&A...649A.177M}. The result that we obtained using \texttt{TLCM} is two times more precise than previous estimates.}

\item{} Spot-crossing events and red noise arising from stellar activity affect the transit parameters. We derived bootstrap accuracy values of 5.5\,\%{} for the transit depth, 3\,minutes for the transit mid-time, and 4.2\,minutes for the transit duration. The derived accuracy is consistent with the observed variation of the transit parameters retrieved from the three individual CHEOPS transits. We also find systematic variations in the transit depth that seem to correlate with the star's brightness, with the deepest transits observed around the photometric minima of the star. These variations are very likely induced by spots that are not spatially located along the transit chord. This will be the subject of a forthcoming separate paper.

\item{}  {We measured a photometric period of $P_{\rm rot}= 4.8367 \pm 0.0006$~d, reflecting the stellar rotation and the brightness variation coming from spots close to the equator. This period is} in a 7:4 commensurability with the orbital period of AU Mic\,b. As it takes 1.75 stellar rotations for AU Mic\,b to complete one orbit around its host star, the transits can only occur at four specific stellar longitudes. We have observations from three of the possible geometrical configurations.

\item{} Because of the 7:4 commensurability between the rotation period of the star and the orbital period of the planet, during the second and third CHEOPS transits AU\,Mic\,b crossed the same stellar surface. We find that the two transits exhibit the same sequence of dark and bright spot-crossing events. We used these events to measure the TTV and found a variation of $\sim$3.9 minutes on a baseline of 33 days. The presence of an outer perturber such as AU\,Mic\,c can account for the observed TTV.
\end{itemize}

\section*{Acknowledgements}

CHEOPS is an ESA mission in partnership with Switzerland with important contributions to the payload and the ground segment from Austria, Belgium, France, Germany, Hungary, Italy, Portugal, Spain, Sweden, and the United Kingdom. The CHEOPS Consortium would like to gratefully acknowledge the support received by all the agencies, offices, universities, and industries involved. Their flexibility and willingness to explore new approaches were essential to the success of this mission.
KGI is the ESA CHEOPS Project Scientist and is responsible for the ESA CHEOPS Guest Observers Programme. She does not participate in, or contribute to, the definition of the Guaranteed Time Programme of the CHEOPS mission through which observations described in this paper have been taken, nor to any aspect of target selection for the programme.
The Belgian participation to CHEOPS has been supported by the Belgian Federal Science Policy Office (BELSPO) in the framework of the PRODEX Program, and by the University of Li{\`e}ge through an ARC grant for Concerted Research Actions financed by the Wallonia-Brussels Federation.

This project has received funding from the European Research Council (ERC) under the European Union’s Horizon 2020 research and innovation programme (project {\sc Four Aces}; grant agreement No 724427).  It has also been carried out in the frame of the National Centre for Competence in Research PlanetS supported by the Swiss National Science Foundation (SNSF). 
YA and MJH  acknowledge  the  support  of  the  Swiss  National Fund  under  grant 200020\_172746. 
SH acknowledges CNES funding through the grant 837319.
ACC and TW acknowledge support from STFC consolidated grant number ST/M001296/1.
PM acknowledges support from STFC, research grant number ST/M001040/1.
ABr was supported by the SNSA.
S.G.S. acknowledge support from FCT through FCT contract nr. CEECIND/00826/2018 and POPH/FSE (EC). {L.D. is an F.R.S.-FNRS Postdoctoral Researcher.}
We acknowledge support from the Spanish Ministry of Science and Innovation and the European Regional Development Fund through grants ESP2016-80435-C2-1-R, ESP2016-80435-C2-2-R, PGC2018-098153-B-C33, PGC2018-098153-B-C31, ESP2017-87676-C5-1-R, MDM-2017-0737 Unidad de Excelencia “Mar\'\i{}a de Maeztu”- Centro de Astrobiología (INTA-CSIC), as well as the support of the Generalitat de Catalunya/CERCA programme. The MOC activities have been supported by the ESA contract No. 4000124370.
S.C.C.B. acknowledges support from FCT through FCT contracts nr. IF/01312/2014/CP1215/CT0004.
DG, MF, SC, XB, and JL acknowledge their roles as ESA-appointed CHEOPS science team members.
This work was supported by FCT - Funda\,c\'ao para a Ci\^encia e a Tecnologia through national funds and by FEDER through COMPETE2020 - Programa Operacional Competitividade e Internacionaliza\,c\'ao by these grants: UID/FIS/04434/2019; UIDB/04434/2020; UIDP/04434/2020; PTDC/FIS-AST/32113/2017 \&{} POCI-01-0145-FEDER- 032113; PTDC/FIS-AST/28953/2017 \&{} POCI-01-0145-FEDER-028953; PTDC/FIS-AST/28987/2017 \&{} POCI-01-0145-FEDER-028987.
B.-O.D. acknowledges support from the Swiss National Science Foundation (PP00P2-190080). 
M.G. is an F.R.S.-FNRS Senior Research Associate.
DG and LMS gratefully acknowledge financial support from the CRT foundation under Grant No. 2018.2323 ``Gaseousor rocky? Unveiling the nature of small worlds''.
MF and CMP gratefully acknowledge the support of the Swedish National Space Agency (DNR 65/19, 174/18).
M.G. is an F.R.S.-FNRS Senior Research Associate.
This work was granted access to the HPC resources of MesoPSL financed by the Region Ile de France and the project Equip@Meso (reference ANR-10-EQPX-29-01) of the programme Investissements d’Avenir supervised by the Agence Nationale pour la Recherche.
This work was also partially supported by a grant from the Simons Foundation (PI Queloz, grant number 327127).
Acknowledges support from the Spanish Ministry of Science and Innovation and the European Regional Development Fund through grant PGC2018-098153-B- C33, as well as the support of the Generalitat de Catalunya/CERCA programme.
This  project  has  been  supported  by  the  Hungarian National Research, Development and Innovation Office (NKFIH) grants GINOP-2.3.2-15-2016-00003, K-119517,  K-125015, the Lend\"ulet Program of the Hungarian Academy of Sciences, project No. LP2018-7/2020, 
the MTA-ELTE Lend\"ulet Milky Way Research Group, Hungary,
and the City of Szombathely under Agreement No.\ 67.177-21/2016.
ZG acknowledges
support from the VEGA grant of the Slovak Academy of Sciences No.
2/0031/18.
L.Bo, G.Pi, I.Pa, G.Sc, G.La, R.R, D.S and V.Na acknowledge the funding support from Italian Space Agency (ASI) regulated by “Accordo ASI-INAF n. 2013-016-R.0 del 9 luglio 2013 e integrazione del 9 luglio 2015”.
G. Bruno is acknowledged for the very fruitful discussions.

{
TESS observations of AU Mic were proposed by the following GO proposals:
G03263 - Plavchan, P;
G03273 - Vega, L;
G03141 - Newton, E;
G03063 - Llama, J;
G03205 - Monsue, T;
G03272 - Burt, J;
G03227 - Davenport, J;
G03228 - Million, Ch;
G03226 - Silverstein, M;
G03202 - Paudel, R.
}

\bibliographystyle{aa}
\bibliography{aumic}

\appendix

\begin{table*}
    \centering
        \caption{Transit times, transit depth, and relative planet radius derived from individual CHEOPS and TESS measurements with \texttt{pyaneti}.  {For comparison, we show the derived transit times of Martioli et al. (2021) converted to BJD, with their published statistical MCMC errors.
        Our errors are precision values derived by parametric bootstraps, the errors in Martioli et al. are MCMC statistical errors describing the core 35\% of the distribution. 
        }}
    \begin{tabular}{|l|l|l||l|}
    \hline
Transit      &  $T_t$ (BJD) [This paper]&  $R_p/R_\star$ [\texttt{pyaneti}] & $T_t$ (BJD) [Martioli et al. (2021)]\\
\hline
CHEOPS \#{}1 & 2459041.2828$\pm 0.0006$ &  0.0541$\pm 0.0002$ & ---\\
CHEOPS \#{}2 & 2459083.5970$\pm 0.0004$ &  0.0520$\pm 0.0002$ & ---\\
CHEOPS \#{}3 & 2459117.4515$\pm 0.0008$ &  0.0515$\pm 0.0002$ & ---\\
TESS S1\#{}1 & 2458330.3911$\pm 0.0008$ & 0.0546$\pm 0.0004$ & 2458330.39046$\pm 0.00016$\\
TESS S1\#{}2 & 2458347.3174$\pm 0.0008$ & 0.0565$\pm 0.0005$ & 2458347.31646$\pm 0.00016$\\
TESS S27\#{}1 & 2459041.2816$\pm 0.0008$ & 0.0517$\pm 0.0003$ & 2459041.28238$\pm 0.00026$\\
TESS S27\#{}2 & 2459049.7457$\pm 0.0008$ & 0.0582$\pm 0.0003$ & 2459049.74538$\pm 0.00026$\\
TESS S27\#{}3 & 2459058.20795$\pm 0.0008$ & 0.0548$\pm 0.0004$ & 2452058.20838$\pm 0.00026$\\
\hline
    \end{tabular}
    \label{table:pyaneti}
\end{table*}

\begin{table}
\caption{Value of roll angle systematics from the \texttt{TLCM} solutions. Values are given in ppm.}
    \centering
    \begin{tabular}{|l|ll|}
    \hline
     & \multicolumn{2}{c|}{CHEOPS \#{}1 } \cr
$\cos_1$, $\sin_1$ &     $79\ +78   / -78$ &
      $208\ +132 / -141 $\\
$\cos_2$, $\sin_2$ & $312\ +104 / -104$ &
      $9\ +104   / -104 $\\
$\cos_3$, $\sin_3$ &      $146\ +69  / -75$ &
      $-257\ +78 / -75 $\\
$\cos_4$, $\sin_4$ &      $0\ +47    / -46$ &
      $-94\ +46  / -46 $\\
   & \multicolumn{2}{c|}{CHEOPS \#{}2} \\
$\cos_1$, $\sin_1$ &      $-54\ +40  / -30 $&
      $30\ +53   / -55 $\\
$\cos_2$, $\sin_2$ &      $47\ +40   / -41 $&
      $29\ +32   / -33 $\\
$\cos_3$, $\sin_3$ &      $69\ +30   / -29 $&
      $41\ +33   / -33 $\\
$\cos_4$, $\sin_4$ &      $-37\ +29  / -29 $&
      $47\ +29   / -29$\\
   & \multicolumn{2}{c|}{CHEOPS \#{}3} \\ 
$\cos_1$, $\sin_1$ &      $92\ +46   / -49 $&
      $-249\ +140/ -151 $ \\
$\cos_2$, $\sin_2$ &     $-400\ +109/ -132 $&
      $-3\ +63   / -59 $ \\
$\cos_3$, $\sin_3$ &     $-19\ +64  / -59 $&
      $112\ +84  / -79 $ \\
$\cos_4$, $\sin_4$ &     $215\ +46  / -46 $&
      $36\ +44   / -48 $ \\
\hline
    \end{tabular}
    \label{tab:roll}
\end{table}

\begin{figure}
    \centering
    \includegraphics[width=0.9\columnwidth]{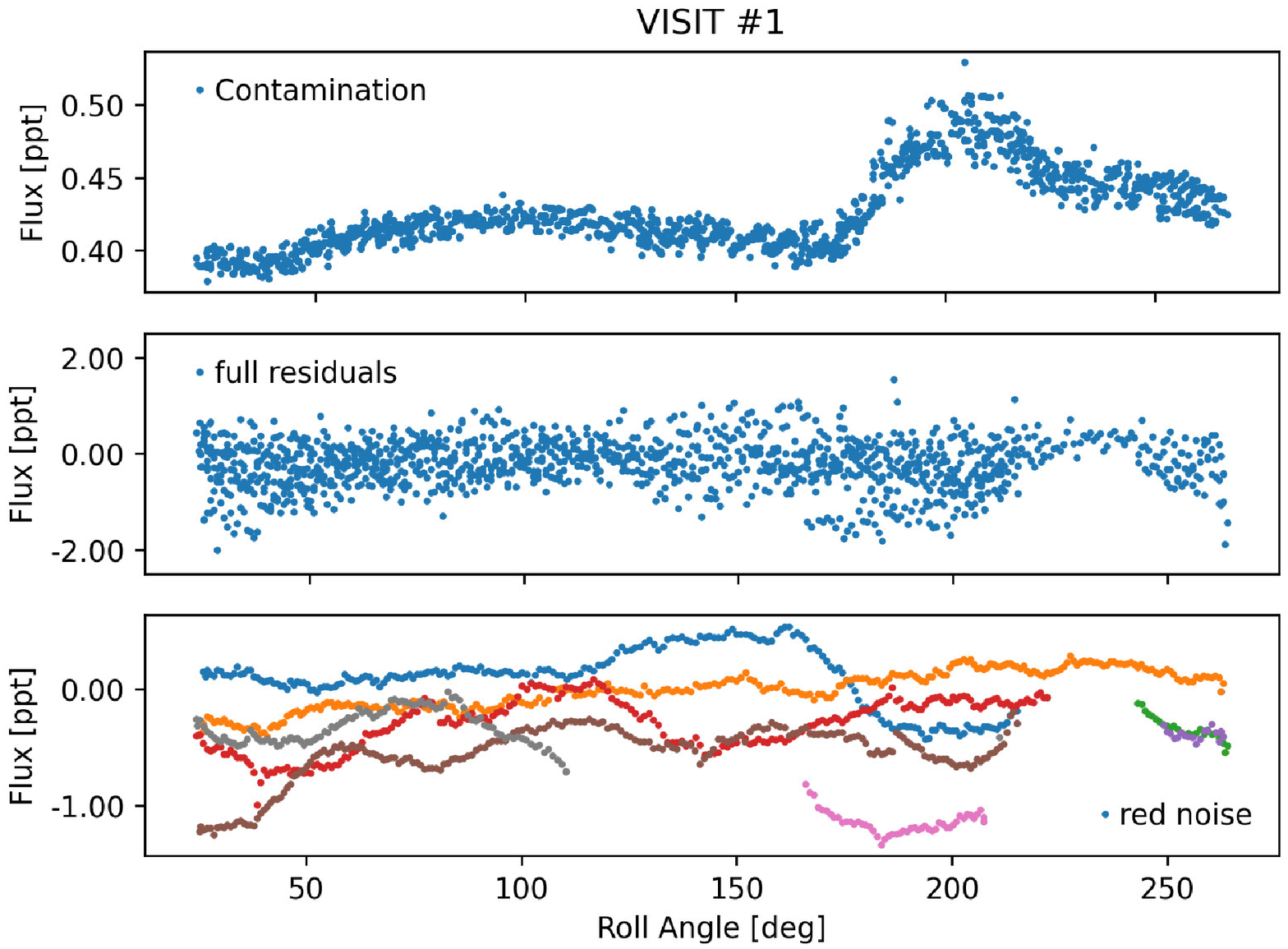}
    \includegraphics[width=0.9\columnwidth]{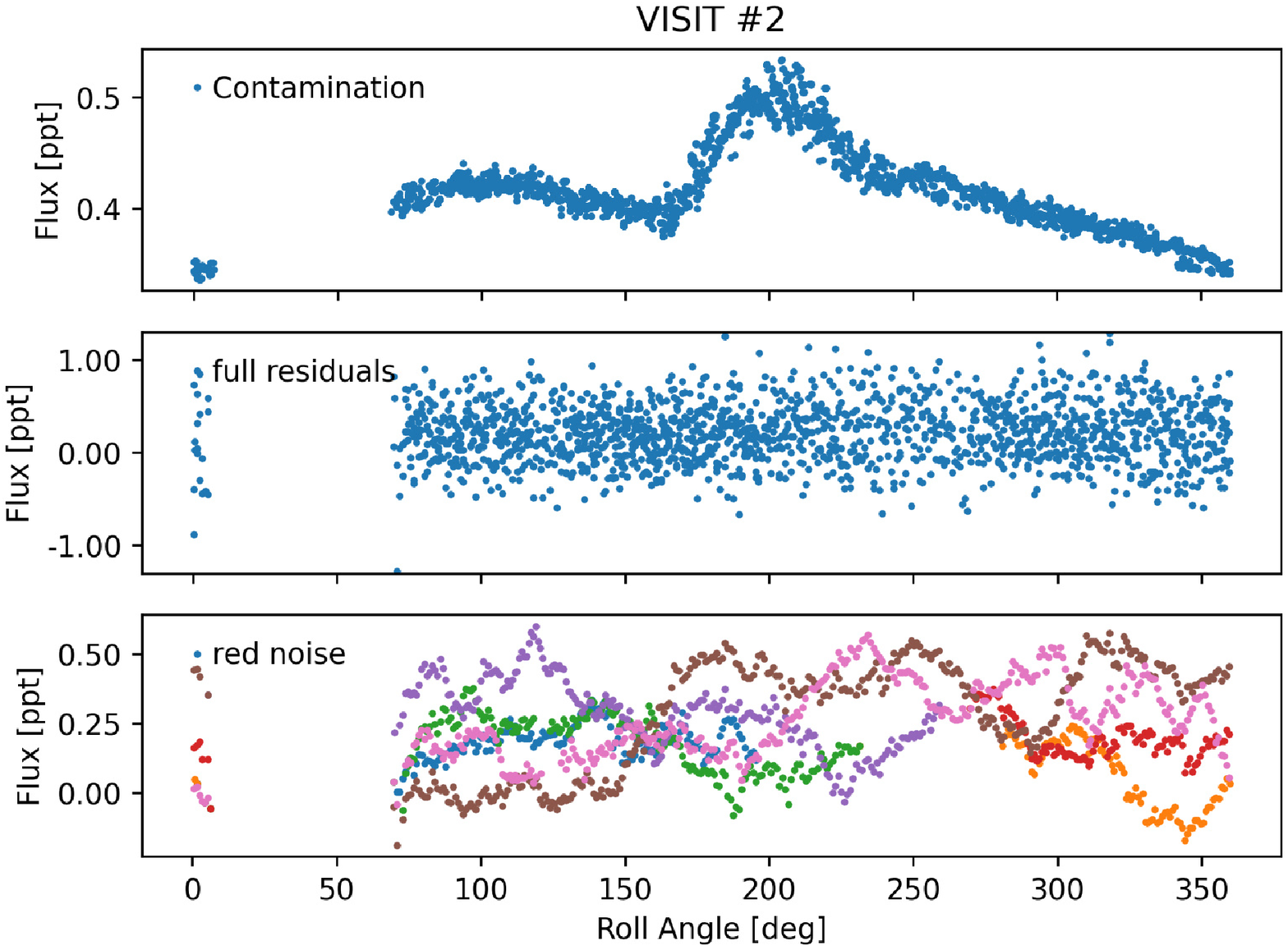}
    \includegraphics[width=0.9\columnwidth]{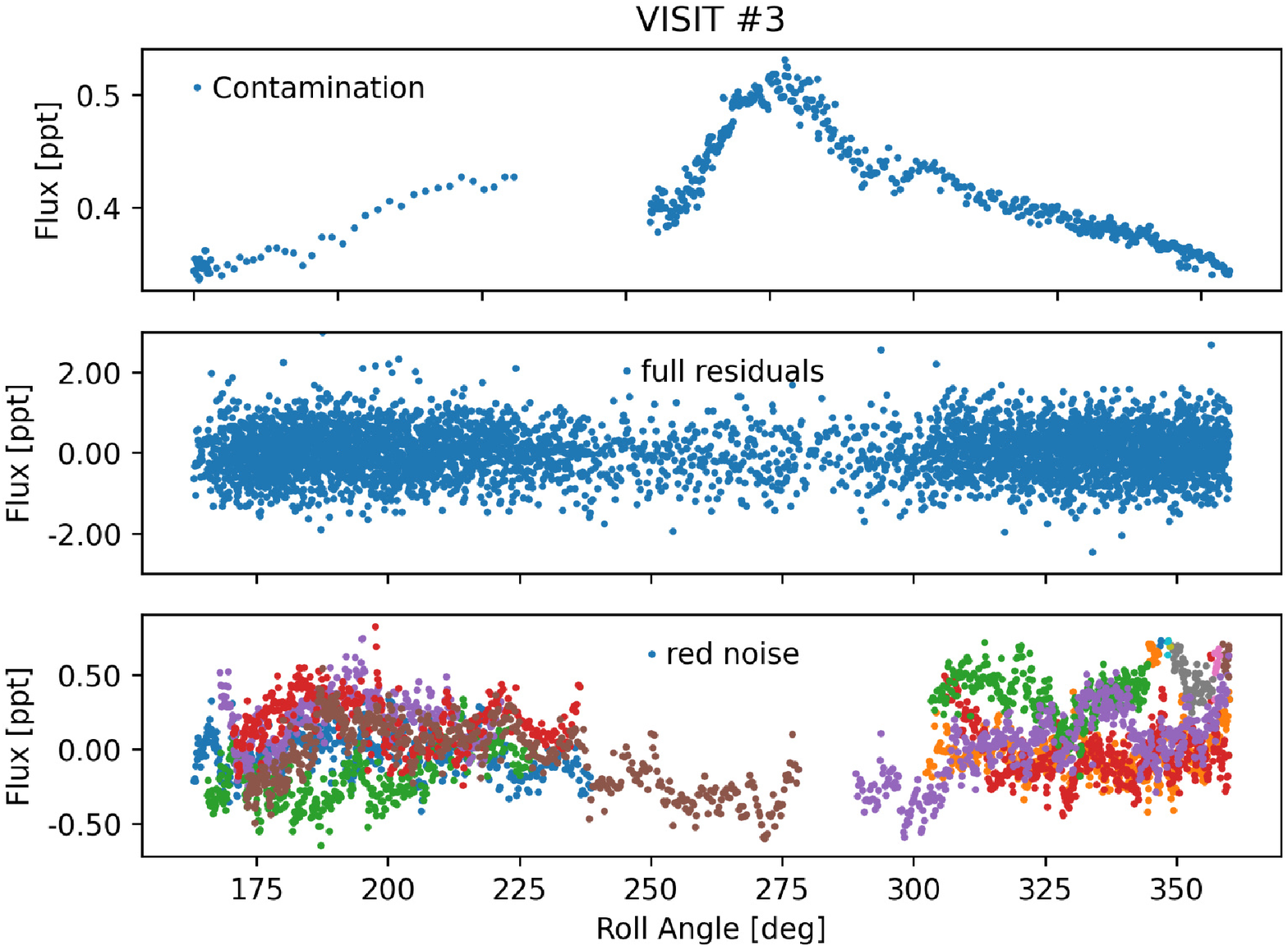}
    \caption{The comparison of light contamination due to the rotation of the telescope (top panels), the local scatter (middle panels) and the shape (bottom panel) of red noise as a function of the roll angle. {The different colours  correspond to different orbits.} The data are well calibrated for roll angle systematics which do not survive significantly in the red noise components of the fit residuals.}
    \label{fig:roll_figure}
\end{figure}

\section{Transit depth and transit times for the individual transits with \texttt{pyaneti}}

{
Table \ref{table:pyaneti} shows the transit times, transit depths, and radius parameters fitted by \texttt{pyaneti} for individual transits involved in the analysis in this paper. We compare the derived minimum times to those of \cite{2020A&A...641L...1M}, and they are general within one error interval.
We also have to note that the errors in Table \ref{table:pyaneti} are derived from massive bootstraps, as described in Section 4.1. and Section 4.3. Bootstrap errors estimate the non-linear biases from noise correlations, and hence, they can be significantly larger than errors from the MCMC statistical fits. In the case of MCMC, the statistical errors show how far our acceptable solutions spread out from the best fit –-the best fit itself can be biased due to noise correlations, too---while the bootstrap tells us how far the best fit can be from the physical parameter set. Therefore, the two error definitions are very different, and cannot be compared directly.}

\section{Determination of roll-angle systematics and the accuracy in planet parameters}

Table~\ref{tab:roll} shows the value of roll-angle systematics from the \texttt{TLCM} solutions. The coefficients belong to a fourth-order polynomial in the form of $\sum \sin_i(i\times \alpha)+cos_i(i\times \alpha)$, where $\alpha$ is the roll angle. 

 Figure \ref{fig:roll_figure}  shows the predicted photometry systematics from the rotation of the telescope (with a period of 97.8 minutes) and the varying scattered light inside the aperture (top panels) for the three CHEOPS visits. These values are the output of the field star modelling algorithm in the CHEOPS DRP, which takes into account the nearby stars and the rotation of the telescope \citep{2020A&A...635A..24H}. This pattern is compared to the scatter (middle panel) and the shape (bottom panel) of the red noise decomposed by the \texttt{TLCM} algorithm from the fit residuals. We can see that the red noise highly exceeds the pattern of scattered light, while the predicted pattern does not repeat in the residuals at all. 

We also performed a correlation analysis using the plotted data. We compared the predicted contamination of scattered light with the measured and separated red noise component at the same time in the actual measurements. We found very low correlation coefficients of $+$0.02 (CHEOPS \#{}1 and CHEOPS \#{}2) and $+$0.03 (CHEOPS \#{}3), while the value of exact $0.0$ is in the 95\% and 98\%{} confidence intervals, respectively. This shows that the calibration and removal of rolling systematics has been done properly, and the red noise pattern we detect is not related to instrument systematics.

\label{lastpage}
\end{document}